\documentclass[a4paper,fleqn]{cas-dc}

\usepackage[authoryear]{natbib}
\usepackage{graphicx}
\usepackage{xcolor}
\usepackage{comment}
\usepackage{graphics}
\usepackage{amsmath}
\usepackage{amssymb}
\usepackage{soul}
\usepackage{yfonts}

\def\CD#1{\textcolor{red}{#1}}

\def\d{\mathrm{d}}

\newtheorem{assumption}{Assumption}

\def\tsc#1{\csdef{#1}{\textsc{\lowercase{#1}}\xspace}}
\tsc{WGM}
\tsc{QE}
\newtheorem{theorem}{Theorem}

\newdefinition{rmk}{Remark}
\newproof{pf}{Proof}
\newproof{pot}{Proof of Theorem \ref{thm}}

\begin{document}
\let\WriteBookmarks\relax
\def\floatpagepagefraction{1}
\def\textpagefraction{.001}

\shorttitle{Recovering nonlinear dynamics from non-uniform observations}    

\shortauthors{C. Donati et al.}

% Option 1
%\title [mode = title]{Recovering nonlinear dynamics from non-uniform observations: A physics-based identification approach with case studies}  

% Option 2
\title [mode = title]{Recovering nonlinear dynamics from non-uniform observations: A physics-based identification approach with practical case studies} 

% Option 3
%\title [mode = title]{Nonlinear system identification from non-uniform observations with case studies of practical interest}  




\author[polito,cnr]{Cesare Donati}[orcid=0000-0003-4368-2753]%} %% Author name

\cormark[1]


\ead{cesare.donati@polito.it}

%\ead[url]{https://www.polito.it/personale?p=cesare.donati}


\affiliation[polito]{organization={Department of Electronics and Telecommunications, Politecnico di Torino},%Department and Organization
            addressline={Corso Duca degli Abruzzi, 24}, 
            city={Torino},
            postcode={10129}, 
            %state={},
            country={Italy}}
\affiliation[cnr]{organization={Cnr-Istituto di Elettronica e di Ingegneria dell'Informazione e delle Telecomunicazioni},%Department and Organization
            addressline={c/o Politecnico di Torino, Corso Duca degli Abruzzi, 24}, 
            city={Torino},
            postcode={10129}, 
            %state={},
            country={Italy}}

\author[cnr]{Martina Mammarella}

\ead{martina.mammarella@cnr.it}

\author[cnr]{Fabrizio Dabbene}

\ead{fabrizio.dabbene@cnr.it}

\author[polito]{Carlo Novara}

\ead{carlo.novara@polito.it}

\author[psu]{Constantino Lagoa}

\ead{cml18@psu.edu}

\affiliation[psu]{organization={School of Electrical Engineering and Computer Science, The Pennsylvania State University},%Department and Organization
            addressline={Electrical Engineering West}, 
            city={University Park},
            postcode={16802}, 
            state={PA},
            country={USA}}

\cortext[1]{Corresponding author}

%\fntext[1]{C. Donati acknowledges support from PRIN project TECHIE “A control and network-based approach for fostering the adoption of new technologies in the ecological transition”, Cod. 2022KPHA24 CUP: D53D23001320006.\\Conflict of interest - none declared}


\begin{abstract}
Uniform and smooth data collection is often infeasible in real-world scenarios. In this paper, we propose an identification framework to effectively handle the so-called non-uniform observations, i.e., data scenarios that include missing measurements, multiple runs, or aggregated observations.
The goal is to provide a general approach for accurately recovering the overall dynamics of possibly nonlinear systems, allowing the capture of the system behavior over time from non-uniform observations.
%
%The proposed approach exploits prior physical knowledge rather than relying solely on black-box models and input-output data. 
The proposed approach exploits prior knowledge by integrating domain-specific, interpretable, physical principles with black-box approximators, proving significant flexibility and adaptability in handling different types of non-uniform measurements, and addressing the limitations of traditional linear and black-box methods.
%First, it is adapted to handle missing measurements and multiple runs. Subsequently, aggregated measurements are addressed by defining an extended system for the identification. 
%
The description of this novel framework is supported by a theoretical study on the effect of non-uniform observations on the accuracy of parameter estimation. Specifically, we demonstrate the existence of upper bounds on the parametric error resulting from missing measurements %and illustrating its dependence on the percentage of missing data and the total number of observations. Additionally, a similar bound is derived for cases with 
and aggregated observations.%, demonstrating how non-uniform observations affect the accuracy of parameter estimation.
Then, the effectiveness of the approach is demonstrated through two case studies. These include a practical application with missing samples, i.e., the identification of a continuous stirred-tank reactor using real data, 
%where our results are compared with existing methodologies in the literature, 
and a simulated Lotka-Volterra system under aggregated observations.
%, a model frequently used to describe population dynamics in ecological studies and interactions in economic contexts.
The results highlight the ability of the framework to robustly estimate the system parameters and to accurately reconstruct the model dynamics despite the availability of non-uniform measurements.
\end{abstract}


\begin{comment}
\begin{highlights}
\item A novel framework for system identification with non-uniform observations is proposed.
\item Missing data, multiple runs, and aggregated measurements handled in a unified approach.
\item Physical system knowledge is leveraged to improve parameter estimation accuracy.
\item Demonstration of theoretical bounds on errors due to missing and aggregated observations.
\item Validated on practical applications, affected by different types of non-uniform observations.
\end{highlights}
\end{comment}

%\nocite{*}

% Keywords

\maketitle

% Main text
\section{Introduction}
Dynamic system identification is a well-established problem in control theory, with a wide range of applications, spanning from engineering to biological systems. On the other hand, when only non-uniform output observations are available \citep{aguero2007system,sleem2024parsimonious}, conventional identification methods, which rely on the assumption that available data are acquired at regular intervals, cannot be effectively employed. In this situation, it becomes crucial to adopt identification techniques capable of integrating non-uniform observations and ensuring consistency despite experimental variability.

\subsection{Case studies and motivations}
The term \textit{non-uniform observations} refers to data scenarios that may include: (i) missing measurements, (ii) multiple runs, i.e., repeated simulations of the system with different initial conditions, or (iii) aggregated outputs, related to either averaged or accumulated observations of the system's outputs over a specified time window rather than individual instantaneous readings.
%i.e., the average or cumulative value 
%
For instance, sensors may fail or operate intermittently and at multi-rate samplings, leading to gaps in the collected data. Another example is represented by data gathered from multiple experimental runs, each starting from a different initial condition or subject to varying external factors \citep{lagoa_missingmeas}.
Moreover, in some specific cases, monitoring the evolution of certain quantities can involve sampling measurements at extended intervals, providing only average values or accumulated information over these periods. This is experienced, for example, in atmospheric or meteorological modeling, where weather stations record average temperature, humidity, or precipitation levels over several hours or days rather than continuously \citep{huntley2010assimilation,amin2024weather}. 
Similarly, in ecological and biological studies, averaged samples are employed to study long-term ecological changes \citep{kidwell2013implications}, or to improve the robustness of statistical analyses and a better understanding of population variability \citep{wangersky1978lotka,nakagawa2011model}.
In economics and finance, data such as gross domestic product (GDP), growth rates, or quarterly earnings are typically collected at extended intervals to provide a broader view of the economic trends and financial health over time (see, e.g., \cite{givoly1982timeliness}). 
In chemical industries, processes like the simulated moving bed involve the collection of samples at extended intervals due to a time-consuming and costly analysis~\citep{grossmann2009system}.

\subsection{State-of-the-art}
In this framework, the majority of the literature focuses on system identification techniques under the assumption of missing data and multiple runs, i.e., measurements are either sporadically absent or completely missing for certain time steps. For example, an expectation maximization-based strategy is presented in \cite{raghavan2006ChemicalMissingMeasurements} for data-driven identification with missing output observations. In this work, 
model parameters and missing observations are simultaneously and iteratively estimated using linear state-space models. In \cite{ARX_missing_measurements}, several reconstruction methods for ARX models with missing measurements were compared, including Kalman filtering, maximum likelihood estimation, and iterative reconstruction. However, all these methods often rely on the exploitation of specific model structures, such as ARX or linear state-space models, which may not generalize well to a large variety of real systems. Furthermore, with this class of approaches, the computational cost for reconstructing missing data becomes increasingly expensive as the amount of missing data grows.

Another class of approaches dealing with missing measurements leverages nuclear norm subspace identification methods, as in \cite{grossmann2009system,liu2013nuclear,varanasi2020nuclear}, where a convex optimization problem is formulated to estimate, in one step, both missing data and model parameters. Despite this class of techniques demonstrates robustness in handling incomplete dataset, their reliance on linear state-space models may jeopardize the model identification if the systems governing equations are highly nonlinear. Alternatively, an expectation-maximization algorithm that employs a particle filter and a particle smoother is employed in \cite{gopaluni2009particle} and \cite{NonlinearSysIDwithMissingObs} for the identification under missing observations of a nonlinear black box model. Analogously, solutions based on black-box neural networks are also proposed in \cite{demeester2020system} and \cite{yuan2023ode} to effectively handle missing observations while identifying a system. Nonetheless, the effectiveness of black-box methods in modeling complex systems is limited by the lack of interpretability \citep{ljung2010perspectives,pillonetto2025deep}. This drawback compromises the possibility to rely on, e.g., prior knowledge of the model structure and physical constraints to compensate for the information lost due to missing measurements. 

On the other hand, to the best of our knowledge, there have been no attempts in the literature to accurately handle the specific problem of aggregated outputs, when available data are temporally averaged or accumulated over fixed, typically non-overlapped, time windows,  leaving a gap in addressing the challenges posed by this type of observation data. However, this aspect needs particular attention, since it introduces an additional level of complexity. Indeed, the aggregation process attenuates short-term fluctuations within the system, thereby concealing the underlying dynamics while leading to misrepresentations of high-frequency components and loss of important information between observation points.

\subsection{Main contribution}
Motivated by the discussion so far, in this paper, we present a novel approach for the identification of nonlinear systems able to handle non-uniform observation conditions, including missing data, multiple runs, and aggregated measurements. Specifically, we enhance the identification performance beyond the available sequence of input-output observations, by integrating the so-called off-white models, i.e., domain-specific physical principles where some parameters have unknown or uncertain numerical values \citep{ljung2010perspectives}, with black-box approximators. Specifically, by leveraging the known physical structure of the system, we ensure that the model remains interpretable, while the black-box term compensates not only the unknown or unmodeled dynamics but also the discrepancies that cannot be resolved due non-uniform observations. In the proposed approach, based on the framework first proposed in \cite{donati2024automatica}, the identification task is formulated as a multi-step optimization problem based on first-order methods. Specifically, we show how, unlike the majority of methods available in the literature (see, e.g., \cite{raghavan2006ChemicalMissingMeasurements,liu2013nuclear}), a minimal modification on either the cost function (for missing measurements and multiple runs) or the estimation model (for aggregated data) allows to extend the method in \cite{donati2024automatica} to the case of non-uniform observations. On the other hand, it is important to highlight that the considered framework relies on first-order methods and the ensuing multi-step problem is in general nonconvex. This clearly leads to the possibility of converging to a sub-optimal solution, and implies an evident sensitivity to initial conditions and black-box approximator choice, necessitating careful tuning of the algorithm’s hyper-parameters and appropriate functions selection to ensure robustness. Despite these limitations, the efficacy of this approach in the case on uniform observations has been already proven in \cite{donati2024automatica}, where the results showed how significantly this novel framework outperformed state-of-the-art methods. Here, we focus on demonstrating how such an informed framework proves particularly advantageous also in scenarios with non-uniform measurements, addressing limitations of traditional linear and black-box methods, and providing a more accurate and interpretable representation of the system dynamics. Moreover, we show how this framework can also be employed for case studies involving temporally averaged or accumulated data over fixed time windows, filling the gap in the literature. 

The investigation on the flexibility of the framework is integrated with a rigorous analysis on the impact of data loss on parameter estimation through the definition of specific theoretical bounds on the estimation error. In particular, we will show the existence of an  upper bound on the parametric error in the case of missing measurements. This bound depends on the percentage of missing data and the total number of observations. Analogously, we demonstrate that a similar bound exists also for aggregated observations, and we show how in this case the accuracy of the identified parameters is influenced by the length of the aggregation window. 

Then, we showcase the effectiveness and robustness of the proposed approach when applied to real-world case studies characterized by challenging dynamics and different types of non-uniform observations. Specifically, we selected two different application fields. The first case study involves the identification with a dataset affected by missing measurements. In this case, real data of a continuous stirred-tank reactor provided by the DaISy database \citep{de1997daisy} -- a collection of real-world datasets frequently used to validate system identification techniques, also including scenarios with missing data (see e.g., \cite{markovsky2013software,varanasi2020nuclear}) -- are utilized.
%are provided by the DaISy database collection \citep{de1997daisy}, including also different scenarios subject to missing observations (see e.g., \cite{markovsky2013software,varanasi2020nuclear}). 
The second case study targets the challenges of working with aggregated data, focusing on the identification of nonlinear ecologic systems. More specifically, we have selected a Lotka-Volterra model, which has been widely used to represent not only the standard predator-prey interactions \citep{wangersky1978lotka} but also the nonlinear dynamics of economic agents or market behaviors \citep{malcai2002theoretical}. In both case studies, the results highlight the efficacy of the proposed framework in accurately reconstructing the system dynamics despite the presence of non-uniform data.

%\CNnote{Summary of contributions?}
%\CD{In summary, the main contributions of this paper are (i) a novel framework for system identification capable of handling non-uniform observation conditions, including missing data, multiple runs, and aggregated measurements, (ii) the derivation of theoretical error bounds to quantify the impact of missing and aggregated observations on parameter estimation, and (iii) the validation of the proposed framework through practical case studies, including the continuous stirred tank reactor and the Lotka-Volterra system.}

%
%To address this problem, we propose a novel method based on the framework proposed in \cite{donati2024automatica}, . The black-box component is designed to compensate for unknown or unmodeled dynamics and the gaps in the available measurements. Moreover, by leveraging the known physical structure of the system, we ensure that the model remains interpretable, while the black-box approximator addresses the discrepancies that cannot be resolved due to missing or averaged data. 

%The proposed framework is first adapted to handle missing measurements and multiple runs. The case of averaged or cumulative measurements is then reformulated as a combination of these scenarios by introducing an extended system model to be identified.

\subsection{Paper outline}
The remainder of this paper is structured as follows. In Section~\ref{sec:recap} the proposed identification framework is briefly introduced, and the extension to multiple runs and missing observations is discussed in Section \ref{sec:nuobs}. Section \ref{sec:cumulative} focuses on the case of aggregated observations, which are treated as a combination of the previous scenarios by the introduction of an extended system model. Results on the two practical case studies affected by non-uniform observations are discussed in Section \ref{sec:numresults} and main conclusions are drawn in {Section~\ref{sec:concl}}.

\vspace{.25cm}
\noindent
\textit{Notation:} Given integers~${a,b\in\mathbb N,\, a\leq b}$, we denote by $[a,b]$ the sequence of integers $\{a,\ldots,b\}$. Given an integer $N>0$ and $N$ vectors ${v_i \in \mathbb R^n}$, $\forall i,\,\forall n$, we denote as $\mathbf{v}_N$ the set of~$N$ vectors $\{v_1,\!\dots,\!v_N\}$. Given a sequence $\mathbf{n}_N$ of $N$ integers and a sequence of vectors in time $v_k \in \mathbb R^n,\,\forall k$, we denote as $\mathbf{v}_{[\mathbf{n}_N]}$ the sequence of vectors at the times contained in $\mathbf{n}_N$, i.e., $\{v_{n_1},\!\dots\!,v_{n_N}\}$. Moreover, given integers ${a,b\in\mathbb N,\, a< b}$, we denote as $\mathbf{v}_{a:b}$ the sequence of vectors from time $a$ to time $b$, i.e., $\{v_a,v_{a+1},\dots,v_{b-1},v_b\}$. The $n\times n$ identity matrix is denoted with $\mathbb{I}_n \in \mathbb{R}^{n, n}$, ${1}_n \in \mathbb{R}^n$ represents a vector of $n$ ones, and ${0}_n \in \mathbb{R}^n$ represents a vector of $n$ zeros. Similarly, $\mathbb{1}_{n,m} \in \mathbb{R}^{n, m}$ and $\mathbb{0}_{n,m} \in \mathbb{R}^{n, m}$ represent an $n \times m$ matrix of all ones and all zeros, respectively. Given a matrix $A\in\mathbb R^{m,n}$, we define its $\ell_p$ norm as $\|A\|_p = \sup_{x\neq0} \frac{\|Ax\|_p}{\|x\|_p}$.

%%%%%%%%%%%%%%%%%%%%%%%%%%%%%%%%%%%%%%%%%%%%%%%%%%%%%%%%%%%
\section{Proposed identification approach}\label{sec:recap} 
In this section, we {briefly} outline the physics-based system identification approach, first proposed in~\cite{donati2024automatica}, which combines an off-white model with black-box approximators. 

We consider a nonlinear, time-invariant system described by the following mathematical model
\begin{equation}
    \begin{aligned}
    \mathcal{S}:\quad&x_{k+1} = f\left({x}_k, {u}_k; \theta\right) + \Delta(x_k,u_k),\\
    &z_k = h\left(x_k; \theta\right),    \end{aligned}
\label{eqn:system}
\end{equation}
where $x \in \mathbb{R}^{n_x}$ is the state vector, $\theta \in \mathbb{R}^{n_\theta}$ is the parameter vector, $u \in \mathbb{R}^{n_u}$ is the input, and $z \in \mathbb{R}^{n_z}$ is the output. Functions $f$ and $h$ are known and assumed to be nonlinear, time-invariant, and continuously differentiable. The term $\Delta$, representing unmodeled dynamics, is unknown.

Given a multi-step input/output sequence of length $T$,
%$\widetilde{\mathbf{u}}_{0:T} = \{\widetilde u_0,\dots, \widetilde u_{T}\}$, $\widetilde u_k = u_k + \eta^u_k$, and the corresponding observations sequence $\widetilde{\mathbf{z}}_{0:T} = \{\widetilde z_0,\dots, \widetilde z_{T}\}$, $\widetilde z_k = z_k + \eta^z_k$,
\begin{equation}
\begin{aligned}
    \widetilde{\mathbf{u}}_{0:T-1} &= \{\widetilde u_0,\dots, \widetilde u_{T-1}\}, &\widetilde u_k = u_k + \eta^u_k,\\
    \widetilde{\mathbf{z}}_{0:T-1} &= \{\widetilde z_0,\dots, \widetilde z_{T-1}\},&\widetilde z_k = z_k + \eta^z_k,
\end{aligned}
\label{eqn:uniform_meas}
\end{equation}
with $\eta^u_k$ and $\eta^z_k$ {being} the input and output measurement noise, 
we seek to estimate the unknown system parameters~$\theta$ and initial conditions~$x_0$, while compensating for the unknown term $\Delta$.
To this aim, we define the following estimation model $\mathcal{M}$ to approximate the system $\mathcal{S}$
\begin{equation}
\begin{aligned}
{\mathcal{M}}\,:\quad\widehat{x}_{k+1} &= {f}(\widehat{x}_{k},  {u}_{k}; \widehat \theta) + {\delta}(\widehat{x}_{k},u_k; \omega),\\
    \widehat {{z}}_{k} &= {h}(\widehat{x}_{k};\widehat \theta),
\label{eqn:bb_extension_v2}
    \end{aligned}
\end{equation}
where $\delta$ is a generic approximator {of the unknown term $\Delta$, e.g., a linear combination of basis functions from a given dictionary, with parameters $\omega \in \mathbb R^{n_\omega}$ to be learned.} 

In the case of uniform observations, the proposed framework considers the prediction error $e_k = \widetilde z_k - \widehat z_k$, and
%Here, $\lambda, \nu, \gamma \in \mathbb R$ are tunable coefficients, while $\mathcal{L}_k$,  $p$, $q$, $r$ are assumed to be twice continuously differentiable functions.
the final goal is to estimate the optimal values of $\theta$, $x_0$, and $\omega$ by solving the following optimization problem
\begin{equation}
    \left( {\theta}^\star, x_0^\star, \omega^\star\right) \doteq \arg \min_{\theta, x_0,\omega}\,\, \mathcal{C}_T(\theta,x_0,\omega; \mathbf{e}_{0:T},\widehat{\mathbf{x}}_{1:T}),
    \label{eqn:optprobl}
\end{equation}
where $\mathcal{C}_T$ is a twice continuously differentiable multi-step cost function, generally defined as
\begin{equation}
\mathcal{C}_T\doteq
    \frac{1}{T}\sum_{k=0}^{T-1} \mathcal{L}_k\left(\theta,x_0,\omega;e_k,\widehat x_k\right),\quad \mathcal{L}_{k}=
        \|\widetilde z_{k}-\widehat z_{k}\|_2^2.
    \label{eqn:final_cost}
\end{equation}
Specifically, the optimal estimation of the unknown system parameters $\theta$, initial conditions  $x_0$, and black-box parameters~$\omega$ is achieved by minimizing $\mathcal{C}_T$ \eqref{eqn:final_cost} over a given horizon~$T$ incorporating terms that account for the squared prediction errors and may also include additional regularization or physics-based penalty terms to enforce constraints derived from the system knowledge (see \cite{donati2024automatica} for further details).
%Notice that \eqref{eqn:final_cost} may include {physics-based} penalties and regularization terms.
As discussed in \cite{donati2023oneshot}, the optimization problem is addressed by using first-order methods. In particular, the gradient of the cost function $\mathcal{C}_T$ is computed at each iteration, and the parameters are updated accordingly, converging to a (potentially sub-optimal) solution. Specifically, we adopt the gradient computation framework from \cite{donati2024gradient}, which leverages automatic differentiation in the context of system identification to find a solution to \eqref{eqn:optprobl} \cite[Algorithm 1]{donati2024gradient}.
The interested reader is referred to \cite{donati2024automatica} for additional details on the core framework.

\section{Missing observations and multiple runs}\label{sec:nuobs}
%As discussed in the introduction, this paper tackles the issue of identifying system parameters from non-uniform measurements. 
In numerous practical scenarios, it is necessary to identify a system from \textit{irregular sampling}, like in the case of missing data, or from multiple runs with unknown initial conditions. In this section, we show how the physics-based approach can be seamlessly extended to include these scenarios.

\subsection{Missing observations}\label{sec:missing_meas}
First, we present the case of\textit{ missing observations}, which arises when only a limited set of output measurements are collected at non-uniform time steps in a given multi-step horizon. Practically, this applies to, e.g., sensor failure, data loss, data cleaning, or limited sampling capabilities. 

To formally define this setup, we consider the uniform sequence of observations in \eqref{eqn:uniform_meas} and the set of \textit{available time steps} $\boldsymbol{\kappa}_N$, i.e., a set of~${N < T}$ ordered integers defined as 
\begin{equation}
    \boldsymbol{\kappa}_N = \{k_1, \dots, k_N\},\, k_j \in [0,T-1],\, \forall j \in [1,N].
    \label{eqn:available_time_steps_vect}
\end{equation}
Accordingly, the sequence of available measurements is defined as follows
\begin{equation}
\begin{aligned}
    %\widetilde{\mathbf{u}}_{[\boldsymbol{\kappa}_N]} &= \{\widetilde u_{k_1},\dots, \widetilde u_{k_N}\},\\
    \widetilde{\mathbf{z}}_{[\boldsymbol{\kappa}_N]} &= \{\widetilde z_{k_1},\dots, \widetilde z_{k_N}\}.
\end{aligned}
\label{eqn:miss_def}
\end{equation}
In this case, the cost function $\mathcal{C}_T$ is defined by selecting only the time steps at which measurements are known to be available, i.e.,
\begin{equation}
    \mathcal{C}_T = \frac{1}{T}\sum_{j=1}^N \mathcal{L}_{k_j},\quad  
    \mathcal{L}_{k_j}=
        \|\widetilde z_{k_j}-\widehat z_{k_j}\|_2^2,
        \label{eqn:kloss_missing_measurements}
\end{equation}
with $k_j$ the $j$-th element of $\boldsymbol{\kappa}_N$. 

%\CNnote{Add some comment on the fact that our approach makes it ``trivial'' but this does not hold for other approaches.}
%We want to remark that the proposed approach requires a minimal adjustment of the cost function to properly handle missing measurements. This flexibility is rarely observed in other standard methods where the missing measurements are, e.g., added as optimization variables or iterative estimation is required, thus possibly increasing the problem complexity \citep{raghavan2006ChemicalMissingMeasurements,liu2013nuclear}.

Despite the simplicity of accounting for missing measurements in the optimization problem, it is crucial to assess their effect on the estimation error in the identified parameters, relying on the assumption on the local identifiability reported in the follows.
\begin{assumption}[local identifiability]\label{ass:identifiability}
The system is locally identifiable according to, e.g., \cite{bellman1970structural} and \cite[Definition 1]{donati2024automatica}. In other words, the Hessian of the loss function evaluated in $\theta^\star$ is always positive definite, i.e.,
$$
H\doteq\left.\frac{\partial^{2}\mathcal{C}_{T}(\theta;\cdot)}{\partial^{2}\theta}\right|_{\theta=\theta^{\star}}\succ0.
$$
\end{assumption}

In the following theorem, which proof is reported in Appendix~\ref{app:Th1}, we demonstrate that there exists an upper bound on the discrepancy between the parameters identified with $N$ available measurements and those obtained with a complete set of $T$ data, and this bound grows proportionally to the square root of the percentage of missing measurements and inversely proportional to $\sqrt{T}$.

\begin{theorem}[error bound with missing measurements]\label{thm:1} Let $\theta_T^\star$ represents the vector of identified parameters obtained by solving problem \eqref{eqn:optprobl} using a complete set of $T$ observations. Similarly, let $\theta_N^\star$ denotes the identified parameters when only $0< N \le T$ observations are available, due to missing data. Define  $p_{\text{miss}} = \frac{T - N}{T}$ as the percentage of missing observations and let Assumption \ref{ass:identifiability} hold. Then, the error between the identified parameters under missing measurements and those from the complete dataset satisfies
\begin{equation}
    \|\theta_T^\star -\theta_N^\star\|_2 \leq \sigma_\xi\frac{1}{\sqrt{T}}\sqrt{p_{\text{miss}}},
    \label{eqn:theo11}
\end{equation}
for some constant $\sigma_\xi \in \mathbb R$.
\end{theorem}

\begin{comment}
The statement of the theorem \eqref{eqn:theo11} is derived considering that $\sigma_\xi = \|\Xi^\top\|_2 \doteq \sigma_{max}(\Xi^\top)$, i.e., the maximum singular value of the matrix $\Xi^\top$, and that for $\gamma_T = \left[\frac1T,\dots,\frac1T\right]^\top$ we have \eqref{eqn:final_cost} equal to \eqref{eqn:new_cost}, while \eqref{eqn:kloss_missing_measurements} is equal to \eqref{eqn:new_cost} considering a vector of coefficient $\gamma_N$\footnote{{Notice that when  $N = 0$, i.e., $p_{\text{miss}}=1$, the system becomes non-identifiable, as $\mathcal{C}_T(\theta, \cdot) = 0$ for all  $\theta$, thereby violating Assumption \ref{ass:identifiability}.}} obtained applying \eqref{eqn:rule} to $\gamma_T$. Thus, we obtain
$$
\|\theta^{\star}(\gamma_T)-\theta^{\star}(\gamma_N)\|_2 = \|\theta_T^\star-\theta_N^\star\|_2 \leq \sigma_\xi\|\gamma_T-\gamma_N\|_2,
$$
for which it is easy to verify that $\|(\gamma_T-\gamma_N)\|_2 =\frac{1}{\sqrt{T}}\sqrt{p_{\text{miss}}}$, concluding the proof.\hfill $\blacksquare$ 
\end{comment}

Theorem \ref{thm:1} establishes a link between the percentage of missing observations, $p_{\text{miss}}$, the multi-step horizon length $T$, and the error in the identified parameters. This bound indicates that the worst-case parametric error increases with the square root of the missing data fraction, highlighting the sensitivity of parameter estimation to the gaps in the data. On the other hand, the factor $\frac1T$ highlights that being the missing data percentage fixed, the datasets collected over shorter horizons (i.e., lower $T$) are inherently more sensitive to missing data, thus leading to a relatively larger error. At the same time, larger datasets effectively help to mitigate the negative impact of missing entries. Moreover, the maximum singular value of the matrix $\Xi$ (see Appendix~\ref{app:Th1}) defines the proportionality constant $\sigma_\xi$. This suggests that systems with certain structural properties (i.e., small $\sigma_\xi$) are more robust to incomplete datasets. 

In the following section, we present a numerical examples that showcases the adherence of the retrieved upper bound with a simulated system subject to missing measurements.

\subsubsection*{Numerical analysis of the upper bound on missing data}
To support the result of Theorem \ref{thm:1}, we demonstrate how missing observations quantitatively affect the parameter identification accuracy by comparing the identification of a second-order linear system with and without missing measurements. The system, presented in \cite{lagoa_missingmeas}, is described by the following transfer function
$$
g(z) = \frac{(\theta_1z + \theta_2)}{(z^2 + \theta_3z + \theta_4)}
$$
and in the canonical companion state-space form as
$$
\begin{aligned}
x_{k+1}&=\begin{bmatrix} -\theta_3 & -\theta_4\\ 1 & 0\end{bmatrix}x_k+\begin{bmatrix} 1\\
     0 \end{bmatrix}u,\\
z_k &= \begin{bmatrix} \theta_1 & \theta_2 \end{bmatrix}x_k,
\end{aligned}
$$
with $\theta_1=0.1037$, $\theta_2=-0.08657$, $\theta_3=-1.78$, $\theta_4=0.9$.
First, we excite the system's step response with a perturbation defined by $\mathcal{N}(0,0.01)$, measured over a horizon ${T=104}$. Then, we use this signal within the proposed approach to identify the system parameters, accounting for different percentages of missing measurements, i.e., from $p_{\text{miss}}= 0.05$ to $p_{\text{miss}}= 0.95$. To achieve this goal, we minimize a cost function of the form \eqref{eqn:kloss_missing_measurements} using a first-order optimization method. The predicted states and outputs are propagated along the horizon $T$, while the gradient is computed via automatic differentiation, relying on the approach outlined in \cite{donati2024gradient,donati2023oneshot}.

\begin{figure}[!tb]
    \centering
    \includegraphics[width=\linewidth]{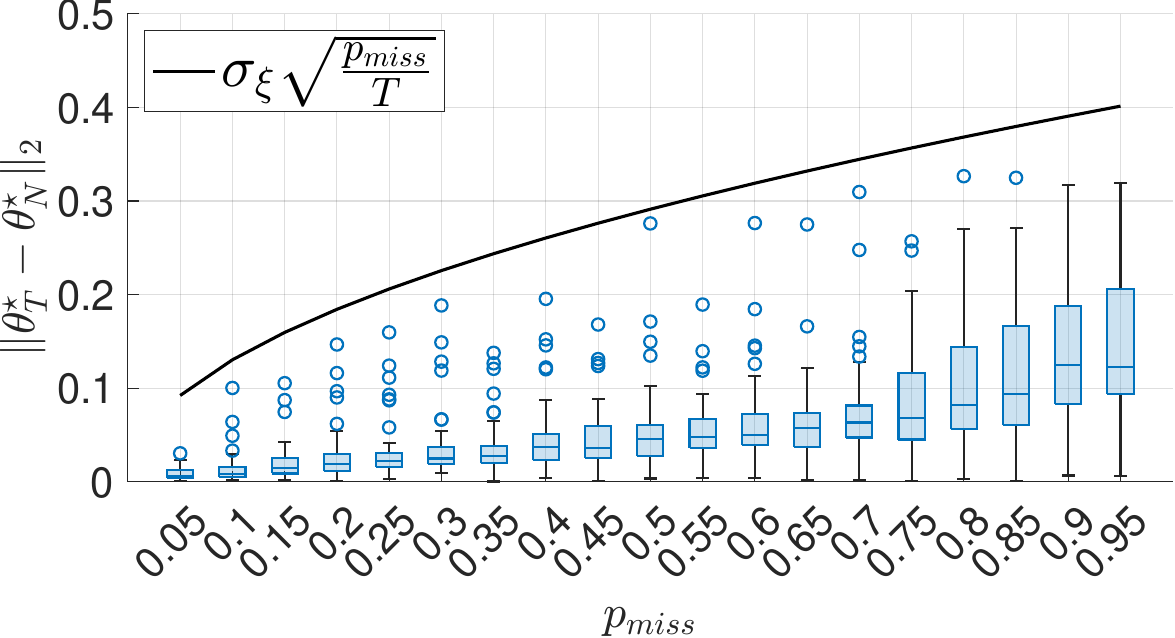}
    \caption{
    Box-plot illustrating the distribution of parameter estimation errors for varying percentages of missing data ($p_{\text{miss}}$). Each box represents the interquartile range (IQR) of errors, with the median shown as the horizontal line, while the whiskers extend the range of errors to values within $1.5\times \text{IQR}$ ($\approx 3\sigma$) beyond the quartiles. Blue dots represent errors that fall outside this range. The black line represents the upper bound trajectory from Theorem \ref{thm:1}}
    \label{fig:thm1_sim}
\end{figure}

Figure~\ref{fig:thm1_sim} shows the relationship between the percentage of missing observations ($p_{\text{miss}}$) and the $\ell_2$ norm of the difference between parameters identified with $p_{\text{miss}} T$ missing data and in the complete-data case (i.e., $p_{\text{miss}}=0$), respectively. Results are collected from $50$ simulations for each percentage of missing data, with each simulation featuring different noise values and initial parameter conditions. We can notice that the error bound scales with the proportion of missing data.
Moreover, when overlapping the upper bound derived in Theorem \ref{thm:1} for $\sigma_\xi=4.2$ to the numerical data, we can observe that this bound well retrace the data behavior, and it appears more conservative at lower values of $p_{\text{miss}}$. Furthermore, we can observe that the shorter boxes reflect estimation errors tightly clustered around the mean, with few outliers approaching the upper bound. This suggests that, for lower percentages of missing data, the identified parameters are more accurate and stable, exhibiting limited variability even in the presence of some data gaps. Conversely, as $p_{\text{miss}}$ increases, the spread of estimation errors around the mean widens, as indicated by the larger boxes. This behavior denotes greater uncertainty and variability in the parameter estimates and it also suggests that a higher amount of missing data makes the identification process less reliable, resulting in a tighter and less conservative upper bound.

\begin{figure}[!tb]
    \centering
    \includegraphics[width=\linewidth]{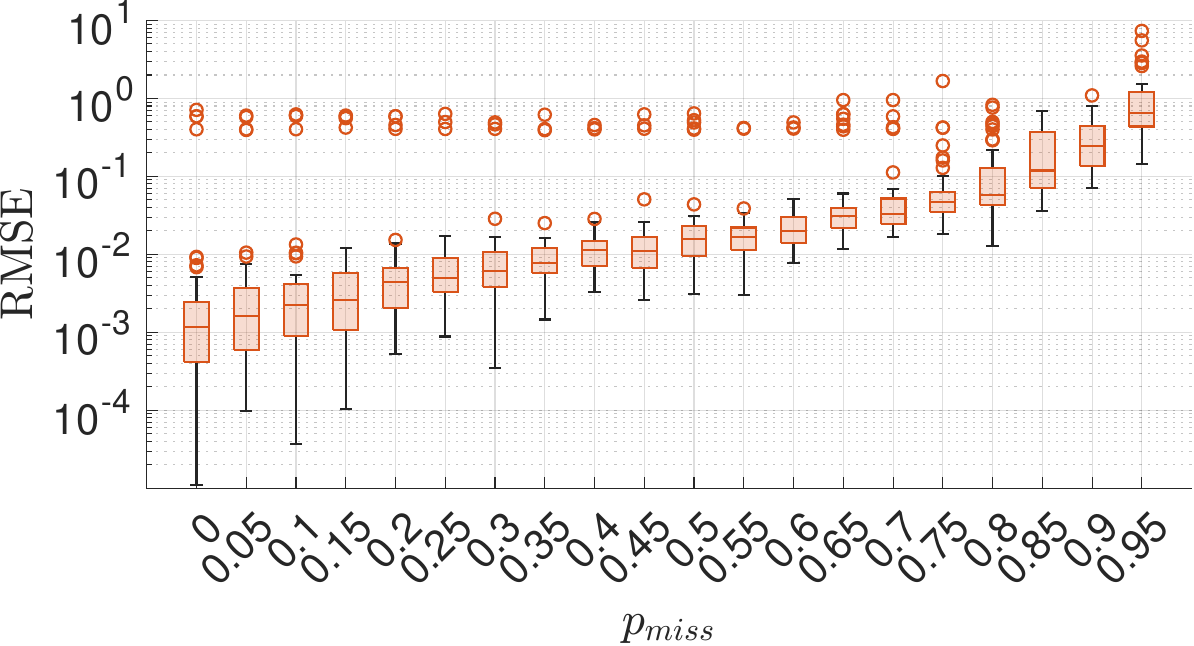}
    \caption{
    Box-plot illustrating the RMSE for varying percentages of missing data (logarithmic scale). }
    \label{fig:thm1_rmse}
\end{figure}

To complete the analysis, Figure \ref{fig:thm1_rmse} shows the root mean square error (RMSE) between predictions and observations for varying percentages of missing data. In this case, the plot reveals, as expected, an increasing RMSE trend with higher $p_{\text{miss}}$ values, thus underscoring the growing discrepancy between predictions and observations as the proportion of missing data rises.

\subsection{Multiple runs}
%On the other hand, 
Now, we focus on the case of \textit{multiple runs}, which occur when the identification process is performed using data from different measured trajectories of the system. This is the case, for example, of repeated experiments under varying initial conditions, different inputs, environmental disturbances, or sensor placements. Additionally, multiple runs are also exploited to identify a more robust model, capable of capturing a wider range of system behaviors and generalizing well to unseen scenarios. Alternatively, multiple runs can be defined from a single trajectory by dividing it into smaller segments (see e.g., the multiple shooting method in \cite{ribeiro2020smoothness}) to obtain specific optimization properties, such as smoothing the cost function and improving numerical stability.
In this setup, we consider $M$ different runs of the system $\mathcal{S}$ \eqref{eqn:system}, starting from $M$ different initial conditions. For each $i$-th experiment, with $i = [1,M]$, the following noise-corrupted sequences of input-output data, each of length\footnote{For simplicity, we assume that each trajectory has the same length $T_r$. However, this non-restrictive assumption can be easily relaxed to accommodate sequences of different lengths.} $T_r$, are available, i.e., 
$$
\begin{aligned}
    \widetilde{\mathbf{u}}^{(i)}_{[0:T_r-1]} &= \{\widetilde u_0^{(i)},\dots, \widetilde u^{(i)}_{T_r-1}\},\\
    \widetilde{\mathbf{z}}^{(i)}_{[0:T_r-1]} &= \{\widetilde z^{(i)}_0,\dots, \widetilde z^{(i)}_{T_r-1}\}.
\end{aligned}
$$
Hence, the cost function $\mathcal{C}_T$ in \eqref{eqn:optprobl} can be redefined as
\begin{equation}
    \mathcal{C}_T = \frac{1}{MT_r}\sum_{i=1}^M \sum_{k=0}^{T_r -1}\mathcal{L}^{(i)}_k,\quad \mathcal{L}^{(i)}_k=\|\widetilde z_k^{(i)}-\widehat z_k^{(i)}\|_2^2.
    \label{eqn:kloss_multiple_runs}
\end{equation}
Here we can notice that in the case of multiple runs the number of decision variables in the optimization problem~\eqref{eqn:optprobl} increases. Indeed, having each run starting from a different initial condition $x_0^{(i)},\,i\in[1,M]$, it follows that the optimization problem \eqref{eqn:optprobl} needs to be minimized with respect to all the initial conditions, i.e., $x_0^{(1)},\dots,x_0^{(M)}$.
%
%
%multiple runs involve the propagation of the prediction of ${\widehat x_k}$ over $T_r$ for each run, starting from $M$ different initial conditions. Thus, 
%
Moreover, it is worth noting that missing measurements and multiple runs can occur simultaneously, as remarked in the following.
\begin{rmk}[multiple runs with missing measurements]\label{rmk:mmmr}
Let us consider $M$ runs with length $T_r$. Then, we define the set of $N<T_r$ available time steps\footnote{Without loss of generality, we use the same $N$ and $T_r$ for each run.} for the $i$-th run as
$$\boldsymbol{\kappa}^{(i)}_{N} = \{k^{(i)}_1, \dots, k^{(i)}_{N}\}, \, k_j \in [0,T_r-1],\,\, \forall j \in [1,N].$$
Accordingly, the available measurements at each run are
$$
\widetilde{\mathbf{z}}^{(i)}_{[\boldsymbol{k}^{(i)}_N]} = \{\widetilde z^{(i)}_{k^{(i)}_1},\dots, \widetilde z^{(i)}_{k^{(i)}_N}\}.
$$
Hence, in the case of multiple runs with missing measurements, the cost function $\mathcal{C}_T$ can be defined by combining \eqref{eqn:kloss_missing_measurements} and \eqref{eqn:kloss_multiple_runs}, i.e., 
    $$
    \mathcal{C}_T = \frac{1}{MT_r}\sum_{i=1}^M \sum_{j=1}^{N }\mathcal{L}^{(i)}_{k_j^{(i)}},\quad \mathcal{L}^{(i)}_{k_j^{(i)}}= 
        \|\widetilde z^{(i)}_{k_j^{(i)}}-\widehat z^{(i)}_{k_j^{(i)}}\|_2^2.
    $$
\end{rmk}

\section{Cumulative or averaged observations}\label{sec:cumulative}
The case of cumulative or averaged observations occurs when over a given time window only gathered information of multiple individual measurements is accessible. This is the case, for example, of monitoring changes of some quantities over extended periods rather than capturing short-term data for practical reasons. In this context, it is important to differentiate two concepts. On one side, we have the so-called running averages, where each measurement is computed as the average of a set of preceding measurements including the current one. In this case, the proposed framework can accommodate running averages as a special case. On the other hand, the primary focus in this paper is on \textit{periodic averaging}, i.e., where averaged observations are considered as the mean of a fixed set of measurements over a specified time window, without, in general, overlapping between consecutive windows.

Unlike the case of multiple runs and missing measurements where a minimal modification of the cost function \eqref{eqn:final_cost} allows to deal with this type of non-uniform observations, in the case of aggregated measurements we need to differently refine the initial reformulation. 
Specifically, let us consider $T$ 
observations and $M$ (possibly consequent) time windows of length $T_r \doteq T/M$ for system \eqref{eqn:system}. Then, the sequence of available measurements can be defined as
%\CDnote{Consider only consequent trajectories?}
\begin{equation}
\begin{gathered}
\widetilde{\mathbf{Z}}_{M} = \{\widetilde Z^{(1)}_{T_r}, \dots, \widetilde Z_{T_r}^{(M)}\}, \quad\widetilde Z^{(i)}_{T_r} = Z^{(i)}_{T_r} + \eta^Z_i\\
Z^{(i)}_{T_r} \doteq \alpha\sum_{k = 0}^{T_r-1} z^{(i)}_k, \quad \forall i \in [1,M],
\end{gathered}
\label{eqn:cum_meas}
\end{equation}
with $\eta^Z_i$ the measurement noise related to the $i$-th cumulative observation, and the parameter $\alpha$ defined according to the type of data, i.e., $\alpha =1$ for cumulative measurements and $\alpha = \frac{1}{T_r}$ when considering averaged measurements.

This non-standard representation of the observations compresses multiple individual measurements into a single data point, concealing short-term dynamics and making it difficult to directly apply standard identification techniques. To address this challenge, we propose an extended system model that reinterprets the problem as one involving both missing measurements and multiple runs,
%to identify while combining the methods used for handling missing data and multiple experiments, 
as detailed next.
%Specifically, referring to the sequence of uniform measurements for multiple runs in \eqref{eqn:multiple_run_meas}, and considering $M$ (possibly consequent) runs, the available cumulative observations are defined as

%

\begin{figure}[!tb]
    \centering
    \includegraphics[trim = {0.5cm 0.5cm 1cm 0.4cm}, clip, width=.9\linewidth]{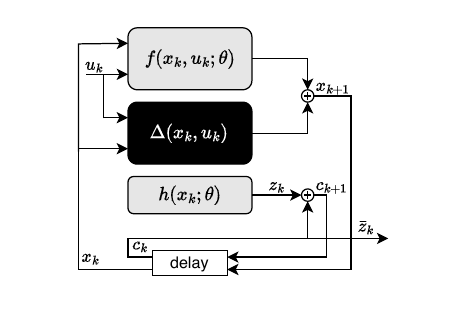}
    \caption{Extended system configuration.}
    \label{fig:scheme}
\end{figure}

%\CDnote{TODO: write this part better. Initial conditions of the new system.}
First, we define the following extended system,
\begin{subequations}
    \begin{align}
    \bar{\mathcal{S}}:\quad& x_{k+1} = f\left({x}_k, {u}_k; \theta\right) + \Delta( x_k,u_k),
    \label{eqn:x_acc_state}\\
    \quad& c_{k+1} =  c_k + h\left( x_k;\theta\right) %+ \eta_k^z,
    \label{eqn:acc_state}\\
    &\bar z_k = \alpha c_k,\label{eqn:acc_obs}
    \end{align}
\label{eqn:system_ext_cumulative}%
\end{subequations}
described by the extended state vector $\bar x = [x_k^\top,c_k^\top]^\top$, with~${c\in \mathbb{R}^{n_z}}$ the cumulative state, which aggregates the outputs $h(x_k; \theta)$ over time, and the new output~${\bar z_k \in \mathbb R^{n_z}}$. The extended system configuration is depicted in Figure \ref{fig:scheme}. 

The following theorem describes how the case of aggregated measurements can be handled by exploiting the tools previously defined for missing measurements and multiple runs, by properly integrating the extended system $\bar{\mathcal{S}}$ \eqref{eqn:system_ext_cumulative}.

\begin{theorem}[systems equivalence]\label{prop:equivalence}
    Let us consider $M$ cumulative (or averaged) observations defined by \eqref{eqn:cum_meas} for the system $\eqref{eqn:system}$, collected from $M$ (possibly consecutive) time windows of length $T_r$. Let $x_0^{(i)}$ be the initial condition of the $i$-th time window.
    Then, let us consider $M$ multiple runs of length $T_r+1$ of the system \eqref{eqn:system_ext_cumulative}, having initial conditions $[x_0^{(i)\top}, 0_{n_z}^\top]^\top$. For each run, let us consider missing measurements, as detailed in Section \ref{sec:missing_meas}, with a vector of available time steps~$\boldsymbol{\kappa_1} = \{T_r\}$. 
    The resulting sequence of available $M$ observations for the extended system in \eqref{eqn:system_ext_cumulative}, i.e., $\bar{\mathbf{z}}^{(i)}_{[\boldsymbol{\kappa}_1]}$ with $i \in [1,M]$,
    corresponds to the cumulative (or averaged) observations defined by \eqref{eqn:cum_meas} for system \eqref{eqn:system}. That is
    \begin{equation}
    \bar{\mathbf{z}}^{(i)}_{[\boldsymbol{\kappa}_1]} = Z^{(i)}_{T_r},\quad \forall i \in [1,M].
    \label{eqn:prop_stat}
    \end{equation}
\end{theorem}
The proof is reported in Appendix~\ref{app:Th2}.

%\CDnote{Proposition:}
%Thus, considering the case of $M$ runs of length $T_r$ with missing measurement with $\boldsymbol{\kappa_1} = \{T_r\}$ for every run $i$ of the system $\mathcal{S}^+$, 
%Hence, cumulative measurements can be managed by accounting for both missing measurements and multiple runs for the system in \eqref{eqn:system_ext_cumulative}. Specifically, for each run, only one measurement is available, representing the accumulation of the outputs within the run, i.e., $\widetilde o_{T_r}$, while $\widetilde o_{0},\dots,\widetilde o_{T_r-1}$ are missing. 

Now, let us define the extended estimation model as
$$
    \begin{aligned}
    \bar {\mathcal{M}}:\quad&\widehat x_{k+1} = f(\widehat{x}_k, {u}_k; \widehat\theta) + \delta(\widehat x_k,u_k;\omega),\\
    \quad&\widehat c_{k+1} = \widehat c_k + h(\widehat x_k;\widehat\theta),\\
    &\widehat{\bar z}_k = \alpha c_k.
    \end{aligned}
$$
From Theorem \ref{prop:equivalence} and Remark \ref{rmk:mmmr}, it follows that, considering $M$ multiple runs and $\boldsymbol{\kappa_1} = \{T_r\}$, the cost function $\mathcal{C}_T$ can be redefined as
\begin{equation}
    \mathcal{C}_T = \frac1M \sum_{i=1}^M \sum_{j=1}^{1 }\mathcal{L}^{(i)}_{k_j} = \frac1M\sum_{i=1}^M \mathcal{L}^{(i)}_{T_r},
    \label{eqn:cost_avg}
\end{equation}
with $\mathcal{L}^{(i)}_{T_r}=\|\widetilde Z_{T_r}^{(i)}-\widehat{\bar z}^{(i)}_{T_r}\|^2_2$ and $\widehat c^{(i)}_0 = 0_{n_z}$ for all $i \in [1,M]$.

The proposed method simplifies the analysis with a common formulation and a more practical implementation by considering the problem of averaged or cumulative measurements analogous to those of missing data and multiple runs, as detailed in Remark \ref{rmk:mmmr}. Indeed, once the data are represented within this framework, the identification process can leverage the previously established algorithms, avoiding non-standard formulations that might arise from directly incorporating data aggregation into a cost function.
%\\\CDnote{begin NEW THM}

Next, we need to assess the contribution of the aggregated observations to the estimation error of the identified parameters. Analogously to the case of missing measurement, in the following theorem we demonstrate the existence of an upper bound on
the discrepancy between the parameters identified with observations aggregated over a window of length $T_r<T$ and those obtained with a complete
set of $T$ can be derived. Moreover, we prove that this upper bound grows proportionally to the square root of ${T_r}$.

\begin{theorem}[error bound with aggregated observations]\label{thm:2}
Let $\theta_T^\star$ be the vector of identified parameters obtained as the solution to the optimization problem \eqref{eqn:optprobl} when a complete set of $T$ observations is available. Similarly, let $\theta_{T_r}^\star$ represent the vector of identified parameters obtained when the observations are aggregated over a window of length $T_r$. Let Assumption \ref{ass:identifiability} hold. Then, the error between the identified parameters under aggregated observations and those obtained from the complete dataset satisfies
\begin{equation}
\|\theta_T^\star - \theta_{T_r}^\star\|_2 \leq L_\theta \beta_{T_r},
\label{eqn:theo1}
\end{equation}
for some constant $L_\theta \in \mathbb{R}$, where $\beta_{T_r}$ is bounded and depends on $\sqrt{T_r}$ as follows
\begin{equation}
\sqrt{T_r} - 1 \leq \beta_{T_r} \leq \sqrt{T_r} + 1.
\label{eqn:theo2}
\end{equation}
\end{theorem}

Theorem \eqref{thm:2} establishes an upper bound on the parametric error when using aggregated observations, showing that the error scales with  the square root of $T_r$, i.e., the length of the aggregation window. This implies that larger aggregation windows (or fewer measurements), can lead to greater deviations in the identified parameters. Similar to the case with missing measurements, this behavior emphasizes the effect of non-uniform observation on the parameter estimation accuracy: while aggregating data may be more practical in some scenarios, the resulting effect can mask short-term dynamics, leading to less accurate identification. 
%\\\CDnote{Add refs to NEW THM in Abstract, Introduction, and Conclusions.}\\
%\\\CDnote{end NEW THM}

\section{Case studies}\label{sec:numresults}
In this section, we present two case studies to demonstrate the efficacy of the proposed framework in handling missing and aggregated observations, respectively. %non-uniform measurements.
%The simulations are run in MATLAB 2024a over an Apple M1 with an 8-core CPU, 8GB of RAM, and a 256 GB SSD unit.
We remark that in both examples the optimization is carried out using a first-order method, propagating the predictions while the gradient is computed through automatic differentiation \citep{donati2024gradient,donati2023oneshot}

\subsection{Identification with missing measurements}
To explore the practical use of the proposed identification method for handling the case of missing data, we consider the continuous stirred-tank reactor (CSTR) described in \cite{morningred1992adaptive}. Specifically, we aim to identify the dynamical models for the CSTR relying on a real dataset with different rates of missing observation extracted from the DaISy benchmarks collection \citep{de1997daisy}. 

\subsubsection{System description}
\begin{figure}
    \centering
    \includegraphics[trim={0 0.55cm 0 0},clip,width=1\linewidth]{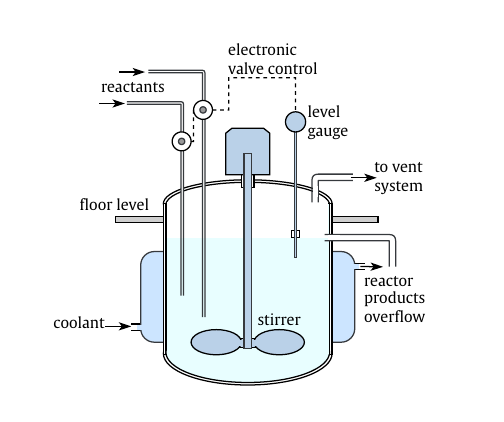}
    \caption{Schematic illustration of a CSTR system.}
    \label{fig:CSTRschema}
\end{figure}

The continuous stirred-tank reactor, depicted in Figure~\ref{fig:CSTRschema}, is governed by an exothermic process with irreversible reaction, where the product concentration is controlled by regulating the coolant flow. 
This system has been widely investigated and it is recognized as a highly challenging benchmark for nonlinear process modeling, optimization, and control (see, e.g., \cite{morningred1992adaptive, lu2014robust, gopaluni2008particle} and references therein). From a system identification perspective, its inherent nonlinear dynamics and sensitivity to operating conditions make it an ideal testbed for validating identification strategies.
%Moreover, this process has been studied in the literature also under conditions of missing data,\cite{gopaluni2008particle, gopaluni2009particle,liu2013nuclear,deng2012identification,yang2018robust,yuan2023ode,demeester2020system}.
Moreover, this process has been studied in the literature also in the case of missing data conditions. Some works, such as \cite{gopaluni2009particle}, \cite{deng2012identification}, and \cite{yang2018robust} focus on the same systems but rely on different datasets, while others, such as \cite{liu2013nuclear}, \cite{yuan2023ode}, and \cite{demeester2020system}, specifically investigate the same dataset used in this paper.

\begin{figure}[!tb]
    \centering
    \includegraphics[width=1\linewidth]{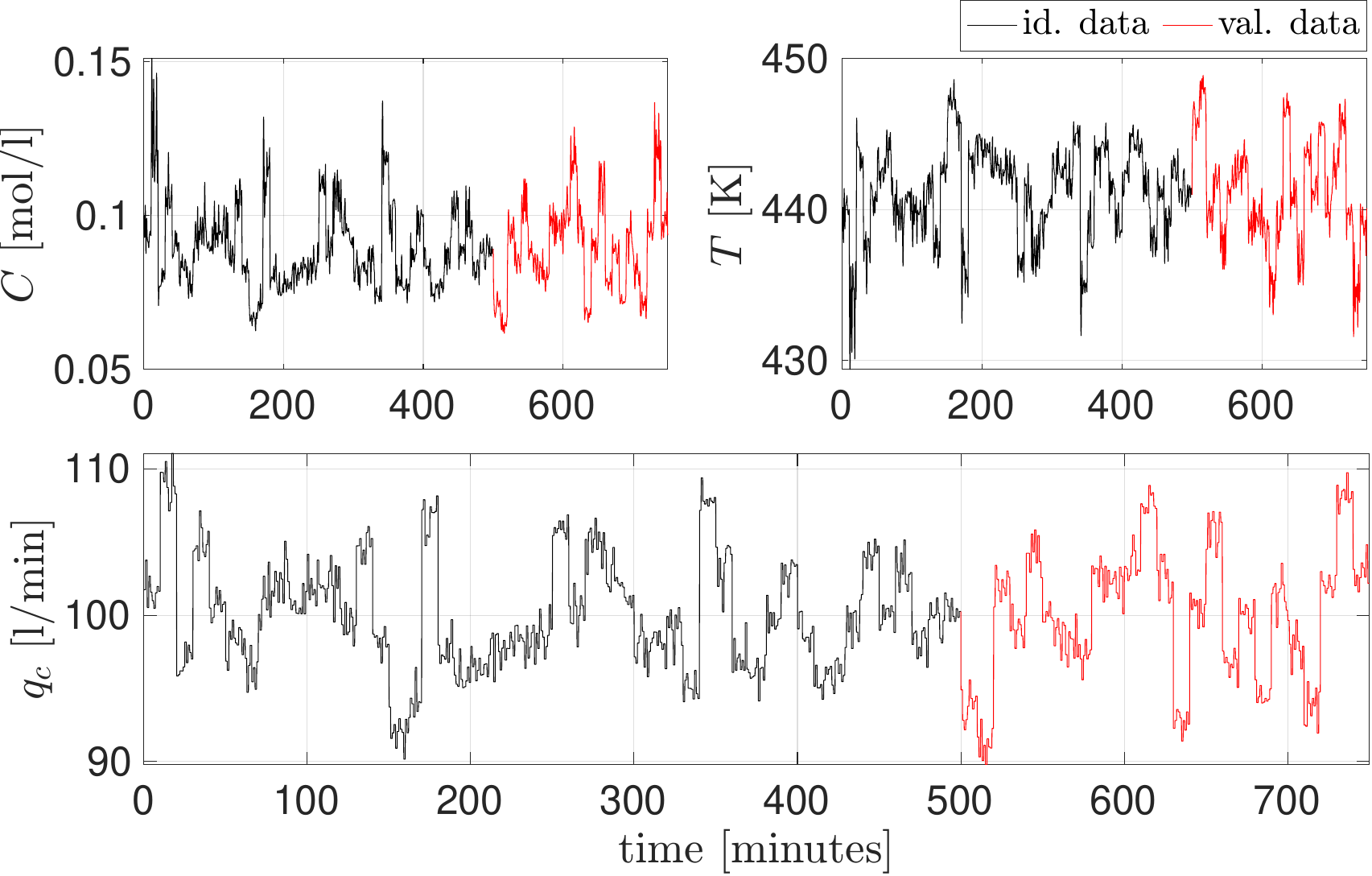}
    \caption{Identification (black) and validation (red) input–output measurements related to the CSTR system.}
    \label{fig:CSTRdata}
\end{figure}

As described in \cite{morningred1992adaptive}, the CSTR system is governed by the following discretized dynamical first-principle equations 
\begin{equation}
\begin{aligned}
    C_{k+1} &= C_k + \Delta t\left[\frac{q}{V}\left(C_0 - C_k\right) - k_0C_ke^{-\frac{E}{RT_k}} \right],\\
    T_{k+1} &= T_k + \Delta t\Big[\frac{q}{V}\left(T_0 - T_k\right) - \frac{\left(-\Delta H\right)k_0}{\rho C_p}C_ke^{-\frac{E}{RT_k}} \\&+ \frac{\rho_cC_{pc}}{\rho C_pV}q_{c,k}\left(1-e^{-\frac{hA}{q_{c,k}\rho C_p}}\right)\left(T_{c0} - T_k\right)\Big],
\end{aligned}
\label{eqn:cstr_eq}
\end{equation}
where the product concentration $C_k$ and the reactor temperature $T_k$ are the state variables, whereas the coolant flow rate $q_{c,k}$ is the input. Moreover, in this system the outputs coincide with the states, i.e., $\widetilde z_k = \left[C_k, T_k\right]^\top$. The goal is to identify the following vector of parameters $\theta=\left[k_0, \frac{\left(-\Delta H\right)k_0}{\rho C_p}, hA\right]^\top$, as in \cite{gopaluni2008particle} and \cite{gopaluni2009particle}. The nominal parameter values used in the simulations and their physical description are reported in Table \ref{tab:cstr_pars}.
\setlength{\tabcolsep}{5pt}
\begin{table}[!tb]
    \centering
    \caption{Nominal CSTR parameter values}
    \begin{tabular}{llcc}
    \hline
         Name & Description & Value & Unit\\
        \hline
        $C_A$ & product concentration & $x_1$ & [mol/l] \\
        $T$ & reactor temperature & $x_2$ & [K]\\
        $q_c$ & coolant flow rate & $u$ &[l/min]\\
        \hline
        $q$ & process flow rate & $100$ &[l/min] \\
        $C_0$ & feed concentration & $1$ & [mol/l] \\
        $T_0$ & feed temperature & $350$ & [K] \\
        $T_{c0}$ & inlet coolant temp & $350$ & [K] \\
        $V$ & CSTR volume & $100$ & [l] \\
        $hA$ & heat transfer term & $7\cdot10^5$ & [cal/min/K]\\
        $k_0$ & reaction rate constant & $7.2\cdot10^{10}$ & [min${^{-1}}$]\\
        $\frac ER$ & activation energy term & $1\cdot10^{4}$ & [K]\\
        $\Delta H$ & heat of reaction & $-2\cdot10^{5}$ & [cal/mol]\\
        $\rho, \rho_c$ & liquid densities & $1\cdot10^{3}$ & g/l\\
        $C_p, C_{pc}$ & specific heats & $1$ & [cal/g/K]\\
        $\Delta t$ & sampling time & $0.1$ & [min]\\
        \hline
    \end{tabular}
    \label{tab:cstr_pars}
\end{table}
The input-output dataset for this process is illustrated in Figure \ref{fig:CSTRdata}. It includes $7500$ samples, $5000$ allocated for the identification task (black line) and the remaining $2500$ (red line) reserved for validation. %For the identification, we used the original dataset and generated different versions with missing data by randomly excluding samples based on different missing rate probabilities.

\subsubsection{Identification results}
The system identification is achieved by minimizing a cost function of the form \eqref{eqn:kloss_missing_measurements} over the identification data, generating different versions of the original dataset with missing measurement rates ranging from $p_{\text{miss}}=0\%$ to $p_{\text{miss}}=75\%$. For each rate of missing data, the results are collected over $200$ simulations, each one employing different initial values for the parameters to be identified. %Then, the optimization is performed at each run using a first-order optimization method propagating the predictions and computing the gradient via automatic differentiation \citep{donati2024gradient,donati2023oneshot}.
Specifically, the estimated parameters are initialized randomly, with each initial value $\widehat{\theta}_{i,0}$ selected within a ball around the nominal parameter value $\theta_i$ and a radius of $30\%$ of $\theta_i$, i.e., $\widehat{\theta}_{i,0}$ is chosen such that $\widehat{\theta}_{i,0} \in [\theta_i - 0.3\theta_i, \theta_i + 0.3\theta_i]$. Furthermore, the states initial conditions to be estimated are initialized at $\widehat x_{0,0} = \widetilde z_0$. 

For the CSTR, the black-box compensation term $\delta(\cdot)$ is introduced into the dynamical model to handle the process nonlinearities and to guarantee adaptation to unmodeled variations in system parameters. Indeed, the CSTR system is typically subject to changes in the environmental and operational conditions, and it may experience fluctuations that the basic physics-based model \eqref{eqn:cstr_eq} alone cannot capture accurately \citep{morningred1992adaptive,deng2012identification}. Hence, in this example the black-box term $\delta$ is defined as a linear combination of sigmoid, softplus, hyperbolic tangent, trigonometric and polynomial functions. Then, a regularization term is introduced in the cost function \eqref{eqn:kloss_missing_measurements} to promote a sparse black-box component. {This is done by minimizing an approximation of the $\ell_1$-norm of the black-box weights $\omega$ (see \cite{donati2024automatica} for further details).}

\begin{figure}[!ht]
    \centering
    \includegraphics[width=0.975\linewidth]{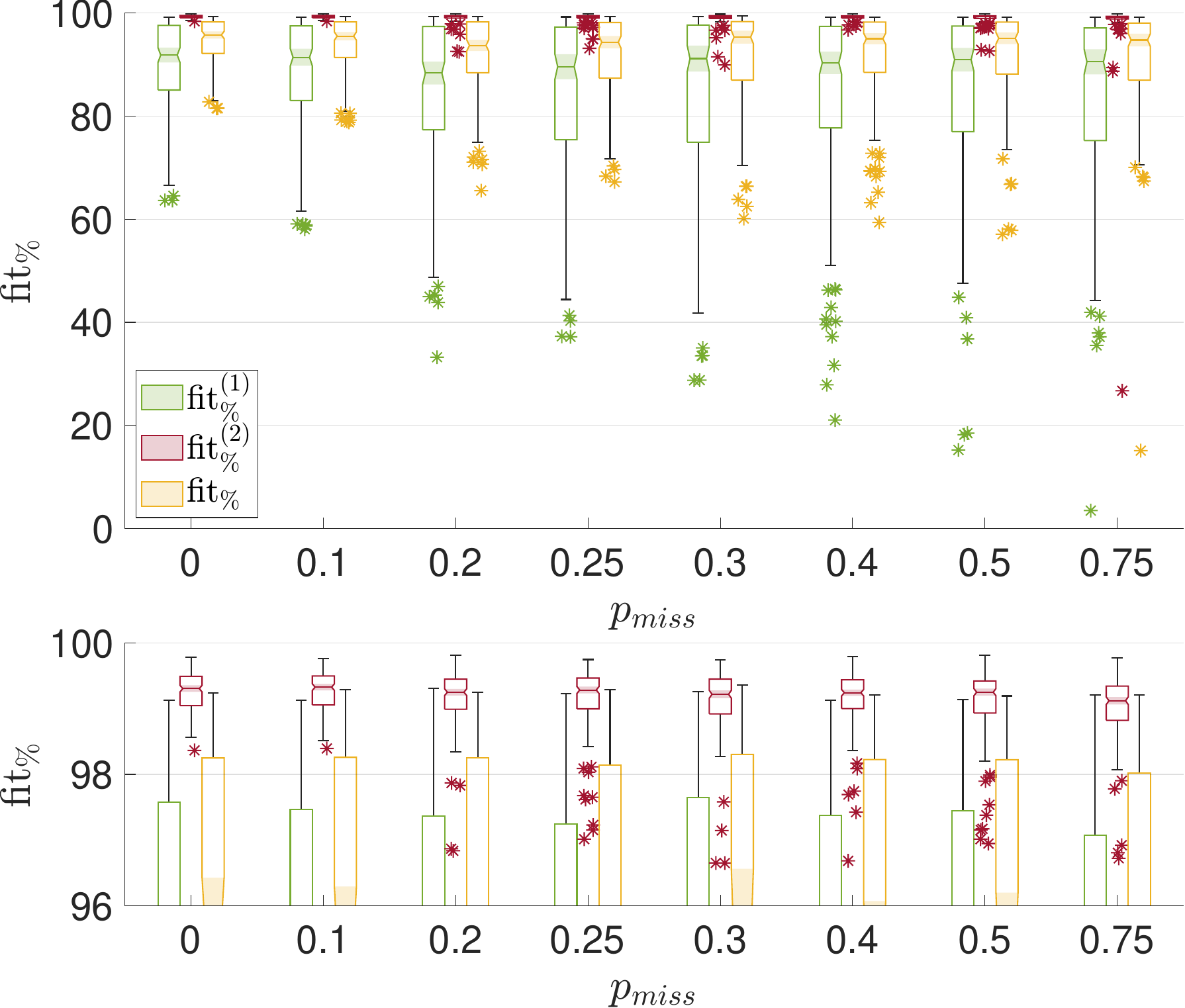}
    \caption{Box-plot illustrating the distribution of the fitness scores for varying percentages of missing data. The bottom plot presents a zoom on the fitness scores for the second output specifically. Each box shows the IQR, with the median as a horizontal line and whiskers extending to 1.5×IQR. Asterisks represent fitness scores that fall outside this range.}
    \label{fig:fit100}
\end{figure}

The results presented next are computed on the validation dataset. First, Figure \ref{fig:fit100} depicts the effect of missing observations on the model fitness scores, which are computed for the $i$-th output as
$$
    \text{fit}^{(i)}_\% = 100\left(1 - \frac{\sum_{k=0}^{T-1}(\widehat z^{(i)}_k - \widetilde{z}^{(i)}_k)^2}{\sum_{k=0}^{T-1} (\widehat z^{(i)}_k - \frac 1T \sum_{k=0}^{T-1} \widetilde{z}^{(i)}_k)^2}\right).
$$ 

\noindent
The box plot highlights the distribution of the model fitness across $200$ simulations for each output and for each level of data loss. The global fitness (yellow boxes) represents the average between the two outputs' fitness, i.e., $\text{fit}^{(1)}_\%$ (green boxes) for the first output and $\text{fit}^{(2)}_\%$ (red boxes) for the second one. Hence, we have ${\text{fit}_\% = (\text{fit}^{(1)}_\%  + \text{fit}^{(2)}_\% )/2}$. 
These results demonstrate the robustness of the proposed approach in the case of in missing data. Indeed, we can observe as increasing the percentage of missing data, the fitness scores remain consistently high, with a gradual decrease observed at higher levels of data loss. Moreover, we have that the narrow interquartile range (IQR) across simulations for lower percentages of missing data implies a low variability in the identified model. On the other hand, the IQR starts increasing for high values of $p_\text{miss}$, reflecting the impact of data loss on the model accuracy and its reliability. Notably, the second output retains better fitness overall, as highlighted in the zoomed section, suggesting a higher resilience of this output to missing observations.

Similarly, Table \ref{tab:fitcomp} presents the global fitness scores, compared to the results obtained in \cite{liu2013nuclear} (Nuc-SId) and \cite{gopaluni2009particle} (PF-NSId) for the same amount of missing data. In this case, the results confirm the ability of the approach in maintaining high global fitness scores across varying levels of missing data, outperforming benchmark results obtained with linear (Nuc-SId) and black-box based (PF-SId) identification methods.

%\begin{comment}
%%% NEW DATA (200 sims)
\begin{table}[!tb]
    \centering
    \caption{Global fitness scores.}
    \begin{tabular}{cccc}
        \hline
         $p_\text{miss}$ & $\text{fit}_\%$ (mean $\pm 1\sigma$) & Nuc-SId & PF-NSId\\
         \hline
         $0$& $94.3\pm4.84$ & $84.7$ & $89.0$ \\
         $10$& $93.7\pm5.47$ & $86.2$ & $88.0$\\
         $20$& $91.9\pm7.49$ & $85.3$ & $87.0$\\
         $25$& $91.8\pm7.61$ & $/$ & $/$\\
         $30$& $91.4\pm8.62$ & $85.4$ &$/$ \\
         $40$& $91.8\pm8.40$ & $85.2$ &$/$ \\
         $50$& $92.1\pm8.28$ & $83.7$ &$/$ \\
         $75$& $91.2\pm9.70$ & $/$ & $/$ \\
         \hline
    \end{tabular}
    \label{tab:fitcomp}
\end{table}
%\end{comment}

\begin{figure*}[!tb]
    \centering
    \includegraphics[width=1\linewidth]{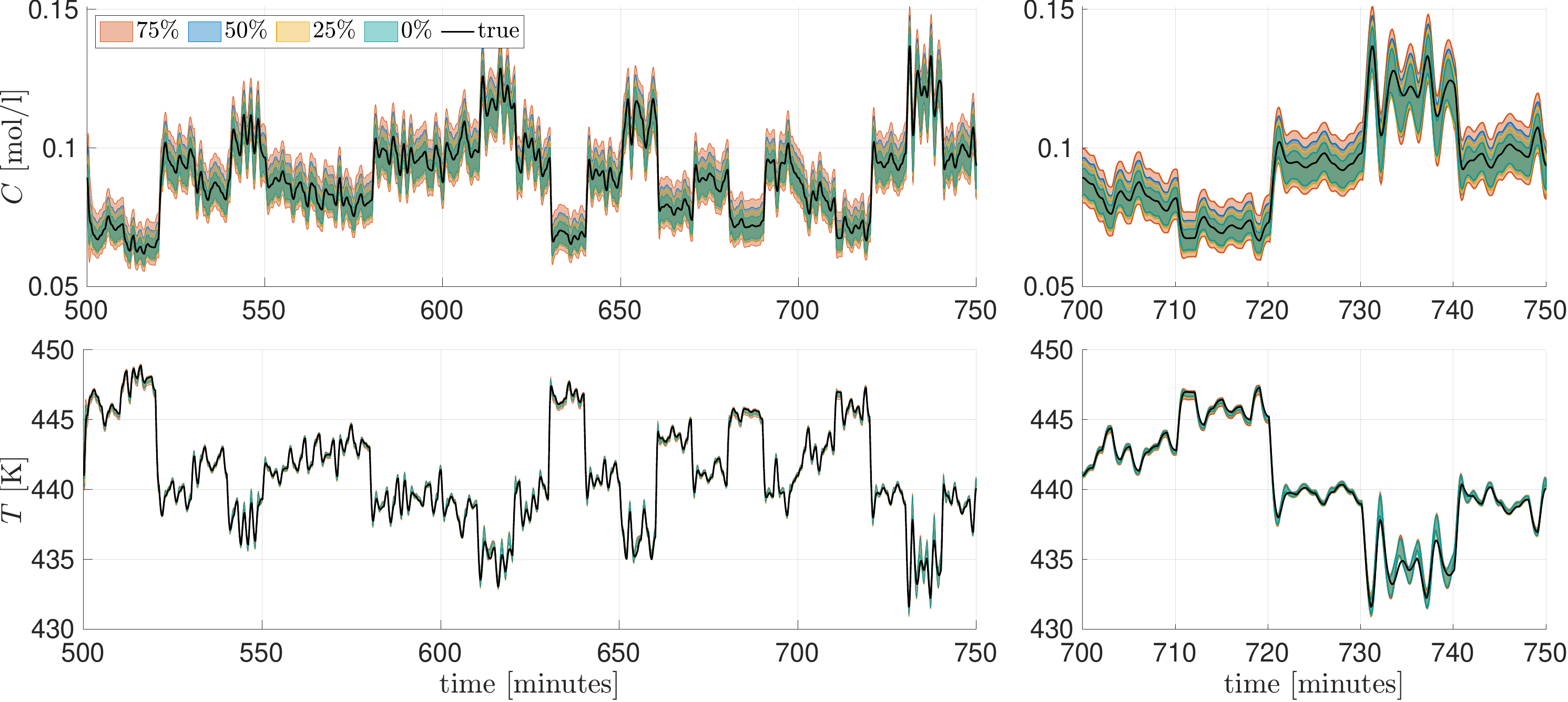}
    \caption{Comparison of true and predicted trajectories for the CSTR system under varying percentages of missing data (indicated in the legend), represented with $\pm1$ standard deviation bands around the mean trajectories. The plots on the right provide a zoomed-in view of the interval $[700, 750]$ minutes to highlight detailed behavior.}
    \label{fig:zh100}
\end{figure*}

Then, the true and predicted trajectories of the CSTR system under four different percentages of missing observations $p_\text{miss}$ are compared in Figure \ref{fig:zh100}, where we have also reported the $\pm 1$ standard deviation bands around the mean trajectories. The results highlight the ability of the proposed method to approximate the system’s dynamics and its inherent robustness to substantial missing data.

Table \ref{tab:rmsetab} collects the RMSE scores for the two outputs, and the reported values demonstrate the ability of the framework to maintain low the prediction errors across varying levels of missing data for both the outputs, while the relatively small standard deviations indicate stable performance, even under significant data loss. Then, Table \ref{tab:rrsetab} shows a comparison with \cite{yuan2023ode} (ODE-RSSM), and \cite{demeester2020system} (NSM-SId) in terms of the relative root squared error (RRSE), which represents the RMSE of the prediction, and it is calculated as~\citep{demeester2020system}
$$
    \text{RRSE} = \frac{1}{n_z}\sum^{n_z}_{i=1}\sqrt{\left(\frac{\sum_{k=0}^{T-1}(\widehat z^{(i)}_k - \widetilde{z}^{(i)}_k)^2}{\sum_{k=0}^{T-1} (\widehat z^{(i)}_k - \frac 1T \sum_{k=0}^{T-1} \widetilde{z}^{(i)}_k)^2}\right)}.
$$
Thus, an RRSE of~$1$ indicates performance equivalent to predicting the mean.
For consistency with \cite{yuan2023ode}, and \cite{demeester2020system} the outputs and predictions were normalized in this phase by subtracting their mean and dividing by their standard deviation before calculation. In this case, the results reported in Table \ref{tab:rrsetab} highlight how the proposed approach is able to provide RRSE values comparable with the ones obtained with the black-box-based benchmark. In particular, better performance are obtained under moderate levels of missing data. Moreover, compared to black-box methods, the proposed approach offers improved interpretability of the identified parameters.

%\begin{comment}
%%% NEW data (200 sims)
\begin{table}[!tb]
    \centering
    \caption{RMSE scores (mean $\pm 1\sigma$).}
    \begin{tabular}{ccc}
        \hline
         $p_\text{miss}$ &$\text{RMSE}_C\times10^{3}$ & $\text{RMSE}_T$\\
         \hline
         $0$&$4.54\pm2.54$& $0.29\pm0.06$ \\
         $10$&$4.88\pm2.85$& $0.29\pm0.06$ \\
         $20$&$5.61\pm3.81$ & $0.31\pm0.09$\\
         $25$&$5.80\pm3.97$ & $0.31\pm0.10$\\
         $30$&$6.09\pm4.76$ & $0.32\pm0.10$ \\
         $40$&$5.94\pm4.96$ & $0.31\pm0.07$\\
         $50$&$5.76\pm4.79$ & $0.31\pm0.09$\\
         $75$&$6.39\pm7.60$ & $0.34\pm0.16$\\
         \hline
    \end{tabular}
    \label{tab:rmsetab}
\end{table}
%\end{comment}
\begin{table}[!tb]
    \centering
    \caption{RRSE scores.}
    \begin{tabular}{cccc}
        \hline
         $p_\text{miss}$ & $\text{RRSE}$ & ODE-RSSM & NSM-SId\\
         & (mean $\pm 1\sigma$) &  & (mean $\pm 1\sigma$)\\
         \hline
         $0$&$0.0787\pm0.0161$ & $0.0659$ & $0.0220\pm0.005$\\
         $10$&  $0.0775\pm0.0152$& $/$& $/$\\
         $20$&  $0.0804\pm0.0173$& $/$& $/$\\
         $25$&  $0.0785\pm0.0161$& $/$& $/$\\
         $30$&  $0.0839\pm0.0192$& $/$& $/$\\
         $40$&  $0.0825\pm0.0166$& $/$& $/$\\
         $50$&  $0.0834\pm0.0187$& $0.1336$ & $0.0920 \pm 0.0140$ \\
         75&   $0.0864\pm0.0249$& $0.2595$ & $/$\\
         \hline
    \end{tabular}
    \label{tab:rrsetab}
\end{table}

Last, the parameters identified for varying percentages of missing data are reported in Table \ref{tab:partab}. Here, the results reflect the level of accuracy and consistency of the parameter estimation achievable under different levels of data loss, demonstrating the ability of the framework to accurately recover interpretable system parameters, maintaining relative reliability across different data loss scenarios. However, it is also important to highlight how the identified parameter may differ from the nominal one, as in the case of $\theta_1$. In the considered case sturdy, this discrepancy may be caused by the variations due environmental and operational conditions, as it commonly happens in real systems.

\begin{table}[!tb]
    \centering
    \caption{Identified parameters (mean $\pm 1\sigma$).}
    \begin{tabular}{cccc}
        \hline
         $p_\text{miss}$ & $\theta_1 \times 10^{-10}$ & $\theta_2 \times 10^{-13}$ & $\theta_3 \times 10^{-5}$ \\
         \hline
         $0$ &$7.64 \pm  0.40$ & $-1.44\pm0.13$ &$7.00\pm0.57$ \\
        $10$&$7.64 \pm  0.37$ &$-1.45\pm0.12$ &$6.93\pm0.50$ \\
        $20$&$7.64 \pm  0.40$ & $-1.44\pm0.13$ & $6.97\pm0.55$ \\
        $25$&$7.64 \pm  0.43$ & $-1.44\pm0.12$ & $6.88\pm0.55$ \\
        $30$& $7.60 \pm  0.43$& $-1.42\pm0.14$ & $6.99\pm0.51$ \\
        $40$&$7.58 \pm  0.40$ & $-1.43\pm0.13$ & $7.01\pm0.52$ \\
        $50$& $7.59 \pm  0.46$& $-1.42\pm0.13$ & $6.91\pm0.55$ \\
        $75$& $7.54 \pm  0.42$ & $-1.41\pm0.14$ &  $6.92\pm0.49$ \\
         \hline
         Nominal & $7.20$ & $-1.44$ & $7.00$ \\
         \hline
    \end{tabular}
    \label{tab:partab}
\end{table}

Summarizing, the application of the proposed approach to the CSTR system in the presence of missing measurements demonstrates its robustness in handling real-world scenarios with substantial missing data. Despite the inherent challenges, the combination of the physics-based model and the black-box component effectively compensates for missing data, accurately identifying the system parameters. Moreover, the obtained results align with those reported in the literature, and a comparison with methods applied to the same benchmark showcases competitive performance particularly under moderate data loss, highlighting the framework’s reliability in practical process modeling and its adaptability to real-world conditions. 

\subsection{Identification with averaged observations}
In this section, we focus on validating the efficacy of the proposed framework in the case of aggregated observations. In particular, we aim to identify a generic Lotka-Volterra model, which has been largely used to describe the dynamics of a variety of real-world systems.

\subsubsection{System description and motivations}
The Lotka-Volterra model consists of a set of nonlinear equations commonly used to describe the dynamics of systems involving different interacting species. Specifically, this model describes how the population of the different species varies over time. This model is well-known for describing the dynamics of biological systems, as the interaction between predator and prey populations~\citep{wangersky1978lotka}. However, its application extends also beyond the ecological domain, as for instance in the economic context, where it is used to represent the wealth of individual investors or the market capitalization of companies~\citep{malcai2002theoretical}.

In both ecological and economical framework, the use of averaged measurements is a common practice. For example, significant biological species fluctuations may occur over the year, and sampling on a specific date or during a short period might yield a distorted view of the population’s typical behavior~\citep{wangersky1978lotka}. On the other hand, high-frequency data might be unavailable in the economic context, and only averaged values over extended periods can be observed~\citep{givoly1982timeliness}. These averages reflect general trends while concealing short-term variations. Consequently, leveraging such aggregated or averaged data within the proposed framework allows for the identification of models that well align with the available data, while contemporary enabling the recovery of the detailed dynamics and an adaptation to the limitations of the available data resolution. 

\subsubsection{Dynamical model} 
\begin{figure}[!b]
    \centering
    \includegraphics[width=0.95\columnwidth]{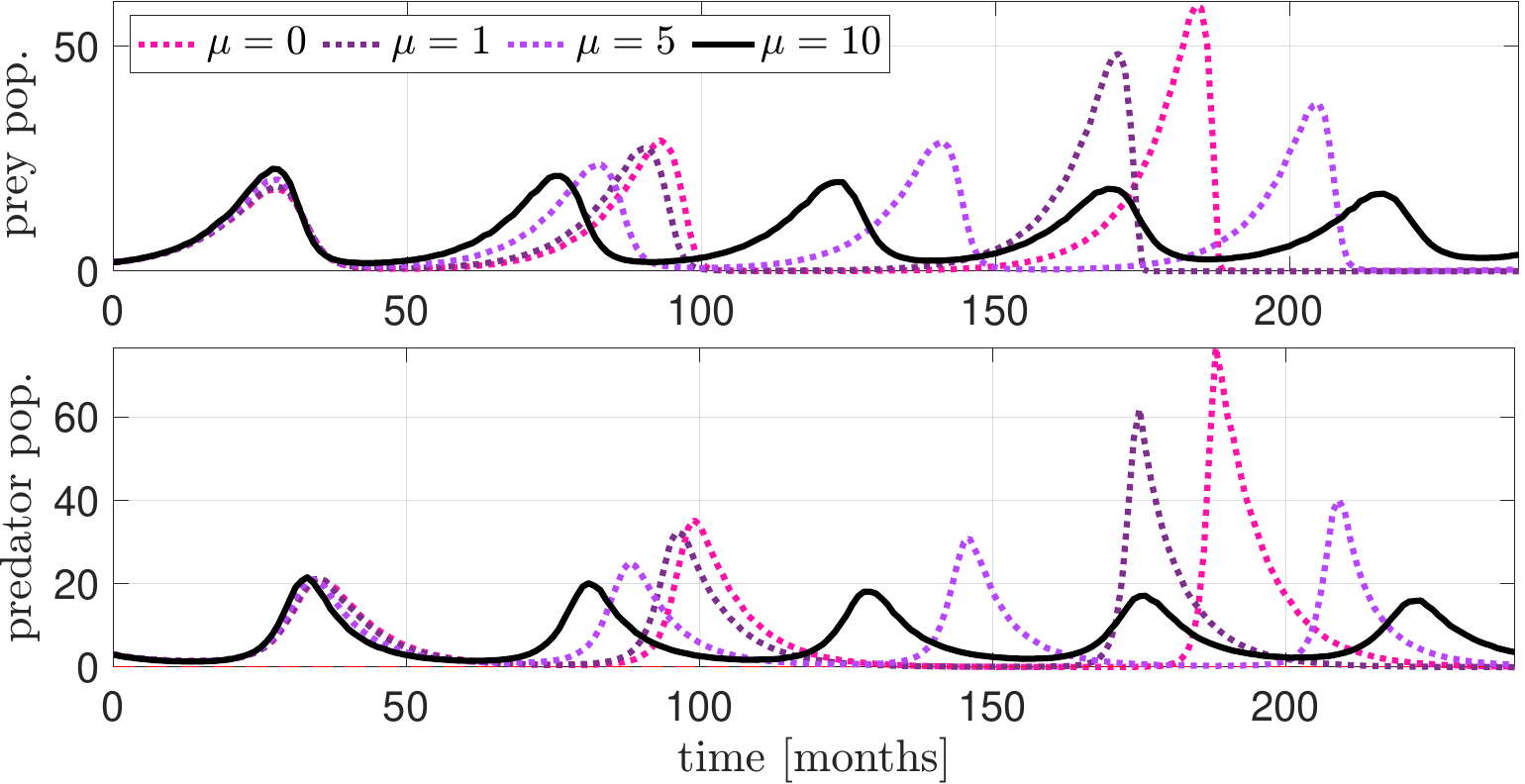}
    \caption{Populations evolution for different values of $\mu$.}
    \label{fig:LV_unm}
\end{figure}

The discretized Lotka-Volterra model with $n_x=2$ states $x_k = [x_{1,k},x_{2,k}]^\top$ and parameters $\theta = [\theta_{1},\theta_{2},\theta_{3},\theta_{4}]^\top$ is given by
\begin{equation}
\begin{aligned}
x_{1,k+1} &= x_{1,k} + \theta_1x_{1,k} - \theta_2x_{1,k}x_{2,k} + \Delta_{1}(x_{k}), \\
x_{2,k+1} &= x_{2,k} - \theta_3x_{2,k} + \theta_4x_{1,k}x_{2,k} + \Delta_2(x_{k}),\\
z_{k} &= x_k,
\end{aligned}
\label{eqn:LVmodel}
\end{equation}
where $x_1$ is the population density of prey, $x_2$ is the population density of the predator, $\theta_1$ and $\theta_2$ are the prey's parameters, describing the maximum per capita growth rate and the effect of predators on the prey growth rate, respectively, $\theta_3$ and $\theta_4$ are the predator's parameters, describing the per capita death rate and the effect of prey on the predator's growth rate, respectively. All parameters are positive and real. On the other hand, $\Delta_1$ and $\Delta_2$ represent unmodeled dynamics that capture external factors affecting the populations beyond the basic predator-prey interaction. 
In the considered case study we have $\theta = [0.13, 0.02, 0.12, 0.02]$, while $\Delta_1$ and $\Delta_2$ are quadratic terms that may represent, e.g., the intraspecific competition within each population, implying that the growth of each population is influenced not only by the interaction between predator and prey but also by the density-dependent effects within each population. In particular, we considered
$$
\begin{aligned}
\Delta_1(x_{k}) &= \mu{10^{-4}} x_{1,k}^2,\quad\Delta_2(x_{k}) = -\mu{5\cdot10^{-4}} 
x_{2,k}^2,
\end{aligned}
$$ 
where $\mu>0$ is a tuning parameter to control the size of the unmodeled terms. In this case study, we consider $\mu=10$. 
Clearly, the unmodeled dynamics, being unknown, cannot be incorporated into the physical model. Instead, it must be compensated by the black-box approximator $\delta$. Notice that, although the unmodeled dynamics seem relatively small, the impact on the population dynamics is relevant. This is highlighted in Figure \ref{fig:LV_unm}, where the evolution of predator and prey populations is represented over a period of 20 years for different values of $\mu$. In this case, it is evident the importance of efficiently compensating for unmodeled dynamics in order to accurately capture the system’s behavior.

%
%Considering \eqref{eqn:bb_extension_v2}, the available physical model used for the identification, i.e., $f(x_k, u_k, \theta)$, is 

\subsubsection{Identification results}
\begin{figure}[!tb]
    \centering
    \includegraphics[width=\linewidth]{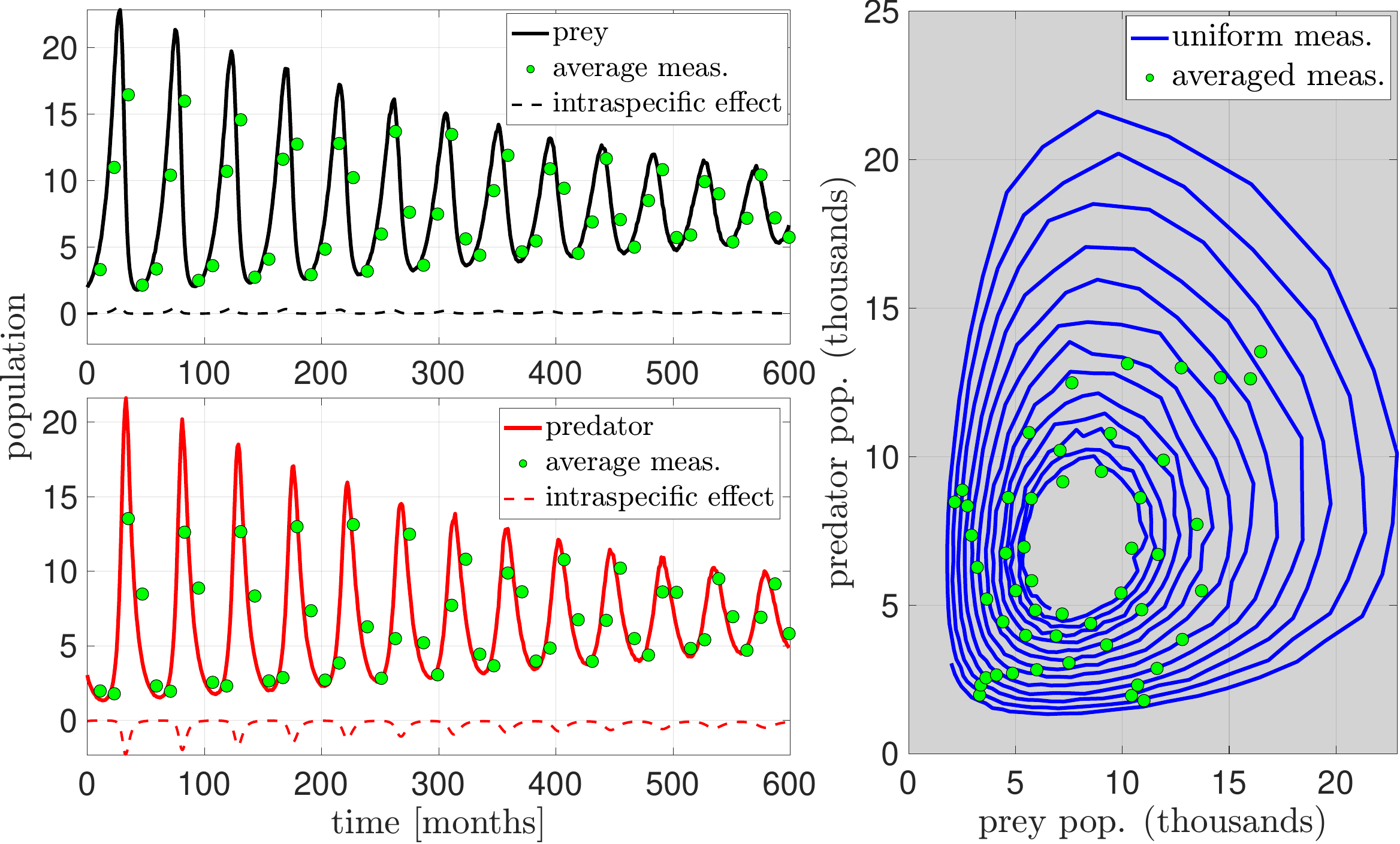}
    \caption{Monthly prey (black lines) and predator (red lines) populations evolutions with yearly average measurements (green circles). Dashed lines represent the isolated effect of the unmodeled dynamics, while the right part of the plot shows the phase plot of the system with uniform measurements (blue lines) and average measurements.}
    \label{fig:LVsys}
\end{figure}

In the proposed example, we simulate the evolution of the predator and prey populations based on the Lotka-Volterra model \eqref{eqn:LVmodel} over 75 years (900 months) with initial condition $x_0=[2,3]^\top$. The data are generated for both populations at \textit{monthly intervals}, capturing the interaction dynamics described by the model. The first $600$ months are used as identification data, while the subsequent $300$ months are used for validation. In the considered scenario, the measures averaged over time windows of $T_r = 12$ months are exploited, leading to $M = 50$ identification measurements for each population. Figure \ref{fig:LVsys} illustrates the monthly population evolution and the yearly average evolution of the prey and predator populations over the entire $50$-year identification period. {This visualization highlights how data averaging captures the overall trend while masking finer details and short-term interactions, which can pose challenges for accurately identifying the system’s underlying dynamics.}

First, we exploit the results of Theorem \ref{prop:equivalence} to identify the system, by employing an extended model of the form 
$$
\begin{aligned}
x_{1,k+1} &= x_{1,k} + \theta_1x_{1,k} - \theta_2x_{1,k}x_{2,k} + \delta_{1}(x_{k}), \\
x_{2,k+1} &= x_{2,k} - \theta_3x_{2,k} + \theta_4x_{1,k}x_{2,k} + \delta_2(x_{k}),\\
c_{1,k+1} &= c_{1,k} + x_{1,k},\\
c_{2,k+1} &= c_{2,k} + x_{2,k},\\
\bar z_{k} &= \frac{1}{T_r}c_k,
\end{aligned}
$$
to estimate the underlying parameters of the predator-prey system from averaged observations. Then, the identification task is performed by minimizing a cost function of the form \eqref{eqn:cost_avg} considering $M$ multiple runs and $\boldsymbol{\kappa}_1 = \{T_r\}$. 

The estimated parameters are initialized randomly as $\widehat \theta_{0} = \theta + \mathcal{N}(0.02, \sigma_{\theta})$, with $\sigma_{\theta}=0.05$. Analogously, the initial conditions of the states are initialized at $\widehat x_{0,0} = \widetilde z_0$. Also in this case, the black-box term $\delta$ is defined as a linear combination of selected basis functions, i.e., sigmoid, softplus, hyperbolic tangent, and trigonometric functions. Additionally, the cost function $\mathcal{C}_T$ incorporates physical penalties and regularization terms to enforce specific properties. Specifically, the positivity of $\widehat{\theta}$ is ensured using an exponential barrier function, while the sparsity of the black-box component $\delta$ is promoted through an $\ell_1$-norm approximation applied to the black-box weights $\omega$ \citep{donati2024automatica}.

\begin{figure}[!tb]
    \centering
    \includegraphics[width=1\linewidth]{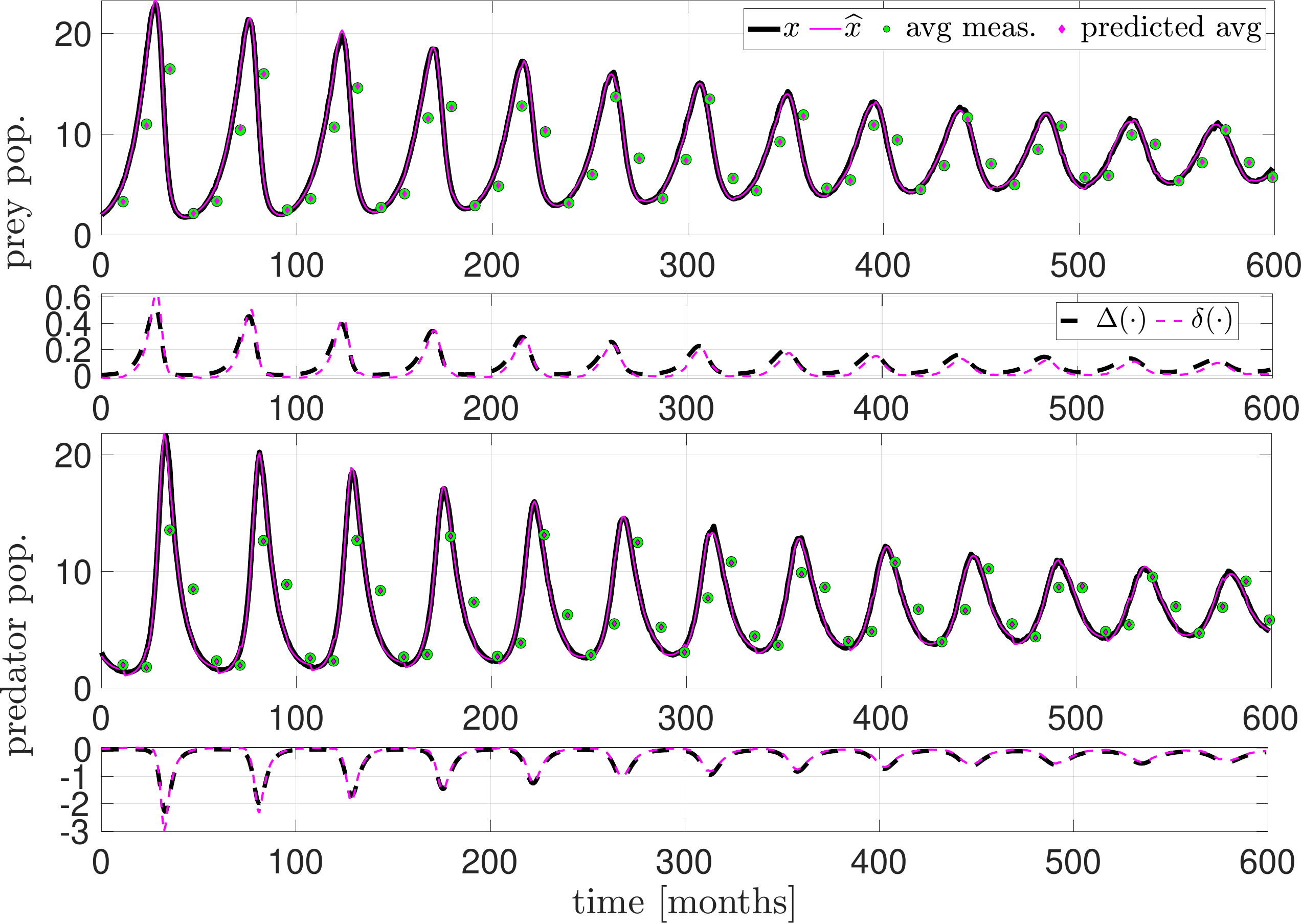}
    \caption{Comparison between the true evolution of the populations (black lines) and the one predicted (purple lines) using the model identified from averaged measurements on the identification data. Green markers represent the averaged measurements, while purple markers represent the reconstructed averages. Purple dashed lines indicate the unmodeled dynamics predicted by the black-box term $\delta(\cdot)$, compared with the true one, $\Delta(\cdot)$ (black dashed line).}
    \label{fig:LVid}
\end{figure}

\begin{figure}[!tb]
    \centering
    \includegraphics[trim={32.5cm 0 0 0}, clip, width=0.9\linewidth]{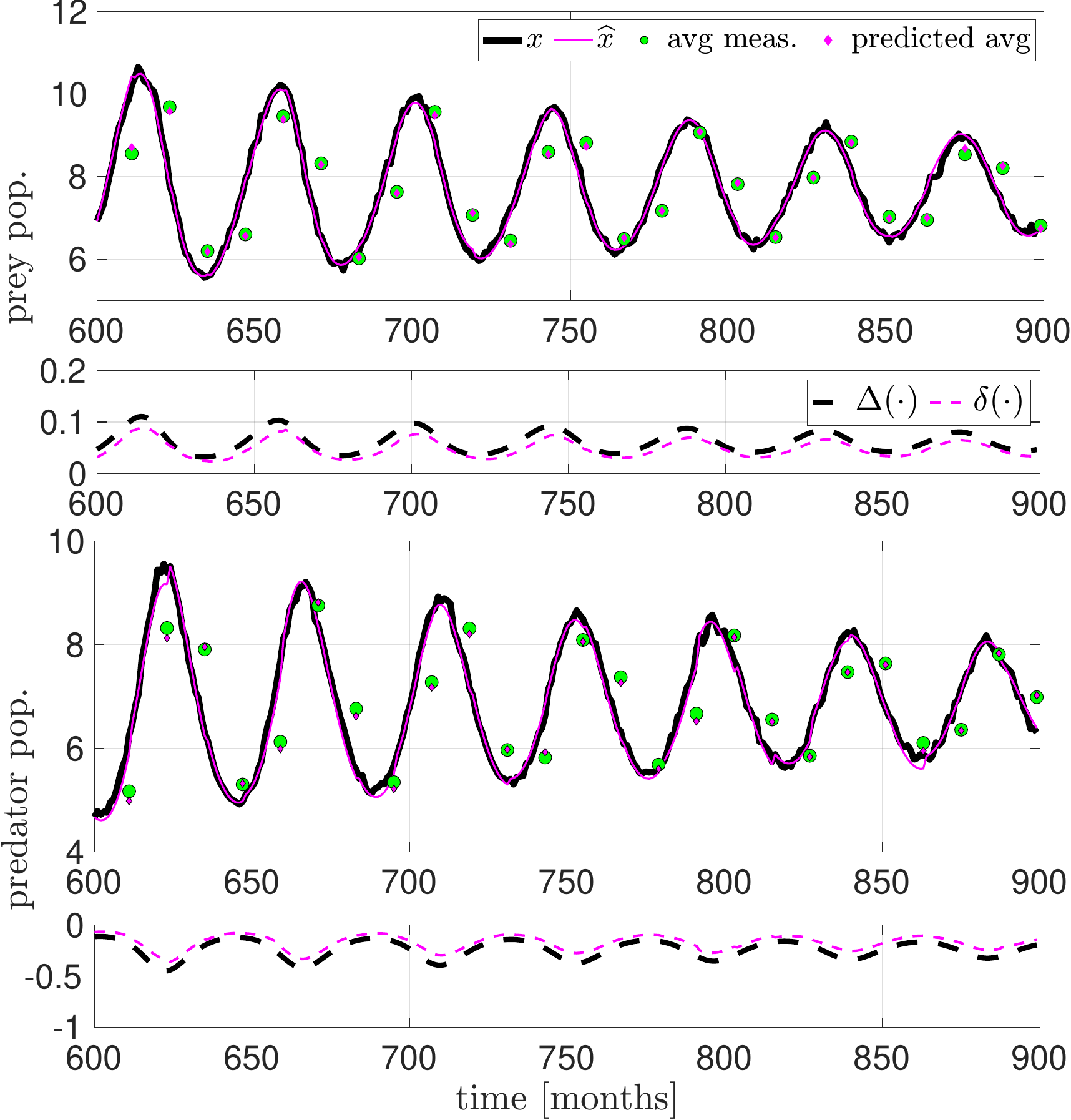}
    \caption{True population evolution (black lines) with the averaged measurements (green markers) and predicted population evolution (purple lines) with the reconstructed averages (purple markers) using the model identified from averaged measurements on the validation data. Black and purple dashed lines represent $\Delta(\cdot)$ and $\delta(\cdot)$, respectively.}
    \label{fig:LVid_val}
\end{figure}

Figure~\ref{fig:LVid} and Figure~\ref{fig:LVid_val} showcase the predictions from the identified model compared with the averaged observations and the population behavior for identification and validation data, respectively. The results highlight how the proposed approach is able to successfully reconstructs the predator and prey dynamics based on the available averaged data and demonstrate the accuracy of the identified model even when facing aggregated measurements.

The accuracy of the method is also reflected in to adherence of the identified parameters and states initial conditions, i.e., $\widehat \theta = [0.1318,\, 0.0204,\, 0.1214,\, 0.0198]^\top$ and $\widehat x_0 = [2.0842, 2.7412]^\top$, with the ground truth values used in the simulation, i.e., $\theta = [0.13, 0.02, 0.12, 0.02]$ and $x_0~=~[2,\,3]^\top$.

Next, we extend the preliminary analysis by considering different sizes of averaging windows. Hence, in addition to the $T_r=12$-month window (yielding $M=50$ averaged data points), we fixed the total amount of data ($T=900$) and the starting value of the estimated parameters and initial condition. Then, we analyze the identification outcomes using larger windows $T_r$, which imply a lower number of available average measurements. Table \ref{tab:table} summarizes the results for each simulated case, including the prediction accuracy measured in terms of the root mean square error\footnote{The RMSE has been computed considering the entire set of observations and predictions over the horizon.}, and parametric error for both $\theta$, for each window size.

\setlength{\tabcolsep}{4.5pt}
\begin{table}[!tb]
    \centering
    \caption{Effect of the averaging window size $T_r$ on identification performance.}
    \begin{tabular}{cccccc}
        \hline
         $T_r$ & $M_\text{tr} / M_\text{val}$ & $\text{RMSE}_\text{tr}$ & $\text{RMSE}_\text{val}$ & $\|\theta-\widehat\theta\|_2$ & $\|x_0-\widehat x_0\|_2$ \\
         \hline
         $12$& $50 / 25$ & $0.2737$& $0.3436$ & $0.0023$& $0.2721$ \\
         $15$& $40 / 20$ & $0.2183$& $0.2846$ & $0.0029$&$0.2047$ \\
         $20$& $30 / 15$ & $0.4742$& $0.5486$ & $0.0123$ & $0.5718$\\
         $24$& $25 / 13$ & $0.8443$& $0.8502$ & $0.0217$ & $0.9885$\\
         %$20$ & $30$& $0.9813$&$0.0052$ & $0.7811$\\
         $40$& $15 / 8$ & $1.4328$& $1.4242$ & $0.0442$ & $1.1534$\\
         $50$& $12 / 6$ & $1.3346$& $1.3539$ & $0.0589$ & $0.9157$\\
         \hline
    \end{tabular}
    \label{tab:table}
\end{table}
%\CDnote{Table I Discussion}
From the reported results, we can observe that, being $T = MT_r$ fixed, an increase in $T_r$ results in fewer available measurements ($M$), which generally leads to less accurate estimates of the parameters and initial conditions, as indicated by increasing error values. Furthermore, a larger $T_r$ not only decreases the data available for identification but also enhances the smoothing effect on short-term dynamics, due to averaging over more values. Consequently, this implies a reduced accuracy, as reflected in the observed trends. Last, it is worth noting that the similar error values observed for $T_r=12$ and $T_r=15$ suggest a range where the performance remains relatively stable, indicating the presence of an ``optimal" averaging window size, beyond which performance begins to degrade. This aspect will be explored in future analysis.

\section{Conclusions}\label{sec:concl}
In this paper, we presented an identification framework able to handle non-uniform observations. The proposed physics-based approach, defined by a proper combination of off-white models and black-box approximators, is designed to address real-world scenarios where measurements may be missing, aggregated, or collected across multiple experimental runs. 
Specifically, for the case of missing measurements and multiple runs, we show how a minimal modification of the cost function allows to handle this class of non-uniform data. On the other hand, for cumulative or averaged measurements, we showed how the identification problem can be handled by relying on an extended system model that reinterprets this problem as one combining missing measurements and multiple runs. In this way, we showcased the highly flexibility of the approach to deal also with non-uniform data.

Moreover, a theoretical analysis has been carried out to assess the effect of non-uniform observations on the estimation accuracy. We demonstrated that for missing measurement there exists an upper bound on the parametric error which depends on the percentage of missing data, the length of the observation horizon, and the resulting error. Analogously, we proved that a similar bound exists also for the case of aggregated observations, highlighting how the length of the aggregation window impacts the accuracy of parameter estimation.

The effectiveness of the proposed approach was demonstrated through two case studies, involving two different types of non-uniform observations. First, we assessed the performance of the approach over a scenario with missing measurements involving the identification of a continuous stirred-tank reactor. Then, we tested the proposed framework in the presence of aggregated data for a Lotka-Volterra system. In both case studies, we demonstrate the ability of the proposed approach to accurately reconstruct system dynamics under non-uniform observations.
% Future works sfruttare robustezza per un framework identification for control. 
%Future works will focus on testing the proposed framework for aggregated observations on real-world data and applications to further validate its robustness and applicability, as well as a more detailed theoretical analysis and test of the averaging window size effect. 

\section*{Conflict of interest statement}
Conflict of interest - none declared

\section*{Acknowledgements}
C. Donati acknowledges support from PRIN project TECHIE “A control and network-based approach for fostering the adoption of new technologies in the ecological transition”, Cod. 2022KPHA24 CUP: D53D23001320006.

\appendix
\section{Proof to Theorem 1}
\label{app:Th1}
Let us consider a cost function of the form 
\begin{equation}
    \mathcal{C}(\theta, \gamma) = \gamma^\top \zeta_T,
    \label{eqn:new_cost}
\end{equation}
where $\gamma \in \mathbb R^T$ is a generic vector of real coefficient and $\zeta_T \in \mathbb R^T$ is a vector containing the squared $\ell_2$-norm of the prediction error for each element , i.e., $\zeta_{T,k} = \|\widetilde z_k - \widehat z_k\|_2^2$. 
Now, since $\mathcal{C}$ depends on $\gamma$, we notice that also a minimizer of this cost is a function of $\gamma$. Hence, denoting with $\theta^\star=[\theta^\star_1,\dots,\theta^\star_{n_\theta}]$ the solution obtained by minimizing a cost function of the type $\eqref{eqn:new_cost}$, we have also that $\theta^{\star}\equiv\theta^{\star}(\gamma)$.
Then, given $\gamma_1\neq\gamma_2$, we define the associated cost functions according to \eqref{eqn:new_cost}, i.e.,
\begin{subequations}
\begin{align}
    \mathcal{C}_1(\theta, \gamma) &= \gamma_1^\top \zeta_T,
    \label{eqn:cost1}\\
    \mathcal{C}_2(\theta, \gamma) &= \gamma_2^\top 
    \zeta_T,
    \label{eqn:cost2}
\end{align}
    %\label{eqn:new_cost2}%
\end{subequations}
Therefore, considering the solutions of \eqref{eqn:cost1} and \eqref{eqn:cost2}, and
%$$\theta_1^{\star} - \theta_2^{\star} \equiv \theta_1^{\star}(\gamma_1) - \theta_2^{\star}(\gamma_2).$$
applying the mean value theorem, for each parameter $\theta^\star_i(\cdot)$ there exists a vector ${\breve\gamma^{(i)}} = (1-a)\gamma_1+a\gamma_2$, with $a \in [0,1]$, such that
\begin{equation} \theta^{\star}_i(\gamma_1)-\theta^{\star}_i(\gamma_2) = {\xi(\breve\gamma^{(i)})}^\top(\gamma_1-\gamma_2),
\label{eq:par_err2}
\end{equation}
with $\xi(\gamma) \doteq \frac{\partial\theta_i^{\star}({\gamma})}{\partial\gamma}\in \mathbb{R}^T$.

Then, to compute $\frac{\partial\theta^{\star}(\cdot)}{\partial\gamma} \in \mathbb R^{n_\theta, T}$ we rely on the implicit differentiation technique. First, we consider that the minimizer $\theta^{\star}$ is a solution of $
\frac{\partial\mathcal{C}_{T}(\theta, \gamma)}{\partial\theta}=0.
$
It follows that
\begin{equation}
\frac{\d}{\d\gamma}\frac{\partial\mathcal{C}_{T}(\theta^\star,\gamma)}{\partial\theta}=0,
\label{eqn:dzero}
\end{equation}
where $\frac{\d}{\d\gamma}$ denotes the Jacobian with respect
to $\gamma$, given by
\begin{equation}
\begin{aligned}
\frac{\d}{\d\gamma}\frac{\partial\mathcal{C}_{T}(\theta^\star,\gamma)}{\partial\theta}%=\frac{\d}{\d\gamma}\frac{\partial\mathcal{C}_{T}(\gamma,\theta^{\star})}{\partial\theta}\\
%\\
 %&
&\!=\!\frac{\partial^{2}\mathcal{C}_{T}(\theta^{\star},\gamma)}{\partial\gamma\partial\theta}+\frac{\partial^{2}\mathcal{C}_{T}(\theta^{\star},\gamma)}{\partial^{2}\theta}\frac{\partial\theta^{\star}(\gamma)}{\partial\gamma}\\
&= G(\gamma) + H(\gamma)\frac{\partial\theta^{\star}(\gamma)}{\partial\gamma},
\end{aligned}
\label{eqn:dcomp}
\end{equation}
where $H$ is the Hessian matrix of the cost function with respect to $\theta$, and $G$ reflects the influence of missing data on the system's behavior.
Therefore, considering \eqref{eqn:dzero}, \eqref{eqn:dcomp}, and knowing that the Hessian $H$ is invertible according to Assumption \ref{ass:identifiability}, we obtain
$$
{\frac{\partial\theta^{\star}(\gamma)}{\partial\gamma}= -H^{-1}G(\gamma).}
$$
This implies that, for the $i$-th parameter $\theta_i^\star$, the vector ${\xi{(\breve\gamma^{(i)})}}^\top$ in \eqref{eq:par_err2} corresponds to the $i$-th row of the matrix $-H^{-1}G(\breve\gamma^{(i)}) \in \mathbb R^{n_\theta,T}$.
Thus, according to \eqref{eq:par_err2}, we have 
$$
\theta^{\star}(\gamma_1)-\theta^{\star}(\gamma_2) = \Xi^\top(\gamma_1-\gamma_2),
$$
 with $\Xi=\left[\xi(\breve\gamma^{(1)}),\dots,\xi(\breve\gamma^{(n_\theta)})\right] \in \mathbb R^{T,n_\theta}$, that yields
\begin{equation}
\begin{aligned}
\|\theta^{\star}(\gamma_1)-\theta^{\star}(\gamma_2)\|_2 &= \|\Xi^\top(\gamma_1-\gamma_2)\|_2 \\&\leq \|\Xi^\top\|_2\|(\gamma_1-\gamma_2)\|_2.
\end{aligned}
\label{eqn:thm1_ineq}
\end{equation}
%for $\gamma_1 \neq \gamma_2$.
%
Let us now consider the case of missing measurements and a set of available time-steps $\boldsymbol{\kappa}_N$ \eqref{eqn:available_time_steps_vect}. 
Specifically, let us define the coefficient vectors in \eqref{eqn:cost1} and \eqref{eqn:cost2} as \begin{subequations}
\begin{align}
    \gamma_1 &= \gamma_T \doteq \left[\frac1T,\dots,\frac1T\right]^\top
    \label{eqn:gammaTdef}\\
    \gamma_2 &= \gamma_N = \left[\gamma_{N,i}\right],\, \gamma_{N,i} \doteq \left\{ \begin{array}{cc}
    0 & \text{if } i \not\in \boldsymbol{\kappa}_N, \\
    \frac{1}{T} & \text{otherwise}. 
\end{array} \right. 
\label{eqn:gammaNdef}
\end{align}
\end{subequations}
%Here, notice that for $\gamma_T$ and $\gamma_N$ the following relations hold, i.e.,
Then, for the coefficient vector $\gamma_T$, the associated cost function $C(\theta, \gamma_T)$ of the form \eqref{eqn:new_cost} is equivalent to the cost function in \eqref{eqn:final_cost}. Indeed, we have
\begin{align}
\label{eq:Cgamma_T}
\mathcal{C}(\theta, \gamma_T) &= \gamma_T^\top \zeta_T = \left[\frac1T,\dots,\frac1T\right]
\left[\begin{array}{c}
    \|\widetilde z_0 - \widehat z_0\|_2^2\\
    \vdots\\
    \|\widetilde z_{T-1} - \widehat z_{T-1}\|_2^2
\end{array}\right] \nonumber\\
&= 
\sum_{k=0}^{T-1} \frac{1}{T}  \|\widetilde z_k - \widehat z_k\|_2^2 = 
\frac{1}{T}\sum_{k=0}^{T-1} \|\widetilde z_k - \widehat z_k\|_2^2.
\end{align}
Analogously, for the coefficient vector $\gamma_N$\footnote{{Notice that when  $N = 0$, i.e., $p_{\text{miss}}=1$, the system becomes non-identifiable, as $\mathcal{C}_T(\theta, \cdot) = 0$ for all  $\theta$, thereby violating Assumption \ref{ass:identifiability}.}} defined in \eqref{eqn:gammaNdef}, the associated cost function $C(\theta, \gamma_N)$ of the form \eqref{eqn:new_cost} is equivalent to one given by \eqref{eqn:kloss_missing_measurements}, i.e.,
\begin{equation}
    \label{eq:Cgamma_N}
\mathcal{C}(\theta, \gamma_N) = \gamma_N^\top \zeta_T = \frac{1}{T}\sum_{j=0}^{N}  \|\widetilde z_{k_j} - \widehat z_{k_j}\|_2^2,
\end{equation}
where $k_j$ is the $j$-th element of $\boldsymbol{\kappa}_N$.
Hence, according to \eqref{eq:Cgamma_T} and \eqref{eq:Cgamma_N}, it follows that \eqref{eqn:thm1_ineq} holds for the solutions $\theta_T^\star$ and $\theta_N^\star$ of \eqref{eqn:final_cost} and \eqref{eqn:kloss_missing_measurements}, respectively. Moreover, defining $\sigma_\xi$ as the maximum singular value of the matrix $\Xi^\top$, i.e., $\sigma_\xi = \|\Xi^\top\|_2 \doteq \sigma_{max}(\Xi^\top)$, we obtain
\begin{equation}
\|\theta^{\star}(\gamma_T)-\theta^{\star}(\gamma_N)\|_2 = \|\theta_T^\star-\theta_N^\star\|_2 \leq \sigma_\xi\|\gamma_T-\gamma_N\|_2.
\label{eqn:thm1_final}
\end{equation}
Then, observing that 
\begin{equation}
\|(\gamma_T-\gamma_N)\|_2 = \sqrt{\frac{T-N}{T^2}}=\frac{1}{\sqrt{T}}\sqrt{p_{\text{miss}}},
\label{eqn:pmiss_gamma}
\end{equation}
and combining \eqref{eqn:thm1_final} with \eqref{eqn:pmiss_gamma}, we yield \eqref{eqn:theo11}, concluding the proof.

\section{Proof to Theorem 2}
\label{app:Th2}
First, we consider the extended system \eqref{eqn:system_ext_cumulative} for a generic run\footnote{In the proof, the superscript $(i)$ is omitted for clarity.} $i$ of length $T_r+1$ with missing measurements defined by $\boldsymbol{\kappa}_1 = {T_r}$. %Moreover, assign to each run\footnote{The superscript $(i)$ is omitted in the proof for clarity.} of \eqref{eqn:system_ext_cumulative} a time window of the cumulative measurements  \eqref{eqn:cum_meas}. 
From \eqref{eqn:miss_def} we have
\begin{equation}
\bar{\mathbf{z}}_{[\boldsymbol{\kappa}_1]} = \{\bar{z}_{k_1}\} = \bar{z}_{T_r}.
\label{eqn:proof0}
\end{equation}
Moreover, from \eqref{eqn:acc_state}-\eqref{eqn:acc_obs}, we obtain 
\begin{equation}
\bar z_{T_r} = \alpha c_{T_r} = \alpha c_{T_r-1} + \alpha h(x_{T_r-1}; \theta).
\label{eqn:proof1}
\end{equation}
Then, iterating \eqref{eqn:acc_state} backward, we obtain
\begin{equation}
\begin{aligned}
c_{T_r-1} &= c_{T_r-2} + h(x_{T_r-2};\theta)\\
&= c_{T_r-3} + h(x_{T_r-3};\theta) + h(x_{T_r-2};\theta)\\
&\vdots\\
&= c_0 + h(x_{0};\theta) + \dots + h(x_{T_r-2};\theta).
\end{aligned}
\label{eqn:proof2}
\end{equation}
Now, from \eqref{eqn:proof1} and \eqref{eqn:proof2} it follows that 
$$
\bar z_{T_r} = \alpha h(x_{0};\theta) + \dots + \alpha h(x_{T_r-1};\theta),
$$
having $c_0 = 0_{n_z}$ by definition.
Moreover, from \eqref{eqn:system} we have ${z_k = h(x_k;\theta)}$, which implies that
\begin{equation}
\bar z_{T_r} = \alpha z_0 + \dots + \alpha z_{T_r-1} = \alpha \sum^{T_r-1}_{k=0} z_k.
\label{eqn:equivalence}
\end{equation}
Finally, considering \eqref{eqn:cum_meas} and \eqref{eqn:proof0}, we have that \eqref{eqn:equivalence} implies \eqref{eqn:prop_stat}, which concludes the proof.

\section{Proof to Theorem 3}
\label{app:Th3}
For simplicity and without loss of generality, let us consider the case of $n_z = 1$. The extension to the general case is straightforward. Given the prediction error $e_k = \widetilde z_k - \widehat z_k$, let us consider the vector
$$\varepsilon = \left[e_0, \dots, e_{T-1} \right]^\top \in \mathbb R^T.$$
Moreover, consider the following cost function $$\mathcal{C}(\Gamma, \theta) = p\|\Gamma \varepsilon\|_2^2,$$
with $p\in \mathbb R$ a generic constant, and $\Gamma \in \mathbb R^{T,T}$ an aggregation matrix of coefficients. 
Given that $\mathcal{C}$ depends on $\Gamma$, any minimizer of the cost function $\mathcal{C}$ will also be a function of $\Gamma$. Therefore, $\theta^\star$, i.e., the solution obtained by minimizing $\mathcal{C}$, can be explicitly expressed as 
\begin{equation}
\theta^{\star}\equiv\theta^{\star}(\Gamma).
\label{eqn:theta_Gamma}
\end{equation}

Thus, considering the cost function $\mathcal{C}(\Gamma,\theta)$ in \eqref{eqn:final_cost} when $T$ uniform measurements are available, we have
\begin{equation}
\mathcal C_T = \frac{1}{T}\sum_{k=0}^{T-1} \|e_k\|_2^2 = \frac{1}{T}\|\Gamma_T \varepsilon\|_2^2 \doteq \mathcal C(\Gamma_T, \theta),
\label{eqn:final_cost_Gamma}
\end{equation}
with $\Gamma_T = \mathbb{I}_T$. Similarly, considering $M$ aggregated measurements with windows length $T_r$, as defined in \eqref{eqn:cum_meas}, and applying \eqref{eqn:equivalence}, we can rewrite \eqref{eqn:cost_avg} as
\begin{equation}
\begin{aligned}\mathcal C_{T} &= 
%\frac{1}{M}\sum_{i=1}^{M} \left\|Z_{T_r}^{(i)} - \frac{1}{T_r}\sum_{k=0}^{T_r-1} \widehat z^{(i)}_k\right\|_2^2\\ 
%&= 
\frac{1}{M}\sum_{i=1}^{M} \left\|\alpha\sum_{k=0}^{T_r-1} \widetilde z^{(i)}_k - \alpha\sum_{k=0}^{T_r-1} \widehat z^{(i)}_k\right\|_2^2\\
%&=\frac{1}{M}\sum_{i=1}^{M} \left\|\frac{1}{T_r}\sum_{k=0}^{T_r-1} (z^{(i)}_k - \widehat z^{(i)}_k)\right\|_2^2\\
&=\frac{\alpha^2}{M}\sum_{i=1}^{M} \left\|\sum_{k=0}^{T_r-1} (\widetilde z^{(i)}_k - \widehat z^{(i)}_k)\right\|_2^2\\
&= \frac{\alpha^2}{M}\left\| \Gamma_{T_r} \varepsilon\right\|_2^2 \doteq \mathcal C(\Gamma_{T_r}, \theta),
\end{aligned}
\label{eqn:cost_avg_Gamma}
\end{equation}
with $\widetilde z_k^{(i)} = z_k^{(i)} 
+ \frac{\eta^Z_i}{T_r}$,
$\Gamma_{T_r}=\left[
    g_1,\dots,
    g_T\right]^\top$, and
\begin{equation}
g_j=\left\{ \begin{array}{ll}
     \left[0_{(j-1)T_r}^\top, 1_{T_r}^\top, 0_{T-jT_r}^\top\right]^\top & \text{if } j\le M,\\
     0_{T} & \text{otherwise.}
\end{array}\right.
\label{eqn:g_def}
\end{equation}

 Thus, being $\mathcal{C}(\Gamma, \theta)$ twice continuously differentiable and the system identifiable according to Assumption \ref{ass:identifiability}, by applying the implicit function theorem \citep{krantz2002implicit} to the gradient function $\nabla_\theta\mathcal{C}(\Gamma, \theta(\Gamma)) = \frac{\partial \mathcal{C}(\Gamma, \theta(\Gamma))}{\partial\theta}$ it follows that $\theta(\Gamma)$ is continuously differentiable, and consequently also Lipschitz continuous, with respect to $\Gamma$. Now, let us consider $\theta_T^\star$ and $\theta_{T_r}^\star$, i.e., the minimizers of \eqref{eqn:final_cost} and \eqref{eqn:cost_avg}, respectively.
 According to \eqref{eqn:theta_Gamma}--\eqref{eqn:cost_avg_Gamma}, we have
 $\theta_T^\star = \theta^\star(\Gamma_T)$ and $\theta_{T_r}^\star = \theta^\star(\Gamma_{T_r})$.
 Therefore, from Lipschitz continuity it follows that
 \begin{equation}
 \label{eqn:betar_theta}
 \|\theta_T^\star - \theta_{T_r}^\star\|_2 \leq L_\theta\|\Gamma_T - \Gamma_{T_r}\|_2,    
 \end{equation}
 for some Lipschitz constant $L_\theta$. 
 Then, defining $\beta_{T_r}$ as the maximum singular value of the matrix $\Gamma_T - \Gamma_{T_r}$, we have
 $$\|\Gamma_T - \Gamma_{T_r}\|_2 = \sigma_{max}(\Gamma_T - \Gamma_{T_r}) = \beta_{T_r},$$
which, combined with \eqref{eqn:betar_theta}, yields \eqref{eqn:theo1}.
 Thus, applying triangle inequality, the following relation holds, i.e.,
 \begin{equation}
 \begin{aligned}
     \beta_{T_r} = \|\Gamma_T - \Gamma_{T_r}\|_2 &\leq \|\Gamma_T\|_2 + \|-\Gamma_{T_r}\|_2 \\&= \|\Gamma_T\|_2 + \|\Gamma_{T_r}\|_2\\&= \sigma_{max}(\Gamma_T) + \sigma_{max}(\Gamma_{T_r}).
     \label{eqn:treq}
 \end{aligned}
 \end{equation}
 Similarly, applying the reverse triangle inequality 
  \begin{equation}
 \begin{aligned}
     \beta_{T_r} = \|\Gamma_T - \Gamma_{T_r}\|_2 &\geq |\|\Gamma_T\|_2 - \|\Gamma_{T_r}\|_2|\\&= |\sigma_{max}(\Gamma_T) - \sigma_{max}(\Gamma_{T_r})|.
     \label{eqn:revtreq}
 \end{aligned}
 \end{equation}
 Here, it is easy to verify that $\sigma_{max}(\Gamma_T) = \sigma_{max}(\mathbb{I}_T) = 1$. Moreover, the following relation holds for all $T_r$, i.e.,
 \begin{equation}
        \sigma_{max}(\Gamma_{T_r}) = \sqrt{T_r}.
        \label{eqn:max_sv_Tr}
\end{equation}
In particular, we have that $\sigma_{max}(\Gamma_{T_r}) = \sqrt{\lambda_{max}(\Gamma_{T_r}\Gamma_{T_r}^\top)}$, where
$$
\Gamma_{T_r}\Gamma_{T_r}^\top\!=\!\left[\begin{array}{c}
        g_1^\top\\
        \vdots\\
        g_T^\top
    \end{array}\right]
        \!\!\Bigl[g_1,\dots,g_T\Bigr]\!=\! \left[\begin{array}{ccc}
        g_1^\top g_1 & \dots & g_1^\top g_T\\
        \vdots & \ddots & \vdots\\
        g_T^\top g_1 & \dots & g_T^\top g_T
    \end{array}\right].
$$
Here, according to \eqref{eqn:g_def}, it is easy to verify that $g_{i,j} = 0$, $\forall i > M$, $\forall j$, and
$$
    g_i^\top g_j = \left\{ \begin{array}{ll}
         1_{T_r}^\top1_{T_r} = T_r &\text{if } i=j\\
         0 &\text{otherwise}
    \end{array}\right.
$$
It follows that 
$$
\Gamma_{T_r}\Gamma_{T_r}^\top = \left[ \begin{array}{ll}
     T_r\mathbb{I}_M & \mathbb0_{M,T-M} \\
     \mathbb0_{T-M,M} & \mathbb0_{T-M,T-M}
\end{array}\right],
$$
is a diagonal matrix, where $\lambda_1 = \dots = \lambda_M = T_r$, $\lambda_{M+1} = \dots = \lambda_T = 0$, and $\sigma_{max}(\Gamma_{T_r}) = \sqrt{T_r}$, proving the statement~\eqref{eqn:max_sv_Tr}.
Thus, from \eqref{eqn:treq}--\eqref{eqn:max_sv_Tr}, we have
$$
|1-\sqrt{T_r}|\leq \beta_{T_r} \leq \sqrt{T_r}+1,
$$
which leads to \eqref{eqn:theo2} having $T_r\geq1$,
%$$
%\sqrt{T_r}-1\leq \beta_{T_r} \leq \sqrt{T_r}+1,
%$$
concluding the proof. Notice that $\beta_{T_r} \approx \sqrt{T_r}$ for large $T_r$.

%%%%%%%%%%%%%%%%%%%%%%%%%%%%%%%%%%%%%%%%%%%%%%%%%%%%%%%%%%%%%%%%%%%%%%%%%%%%%%%%

%\subsection*{Declaration of competing interest}The authors declare that they have no known competing financial interests or personal relationships that could have appeared to influence the work reported in this paper.

\printcredits

\bibliographystyle{cas-model2-names}

\bibliography{main.bib}



\end{document}